\documentclass[aps,prd,preprintnumbers,superscriptaddress,showpacs,amsmath,amssymb]{revtex4-1}

\usepackage{graphicx}
\usepackage{hyperref}

\newcommand{\BDlnu}{\ensuremath{B\to D\ell\nu_\ell}}
\newcommand{\vcb}{\ensuremath{V_{cb}}}
\newcommand{\Vcb}{\ensuremath{|\vcb|}}
\newcommand{\invfb}{\ensuremath{\mbox{~fb}^{-1}}}
\newcommand{\BXlnu}{\ensuremath{B \to X \ell \nu_{\ell}}}
\newcommand{\BXulnu}{\ensuremath{B \to X_u \ell \nu_{\ell}}}
\newcommand{\BXclnu}{\ensuremath{B \to X_c \ell \nu_{\ell}}}
\newcommand{\BDstarlnu}{\ensuremath{B\to D^* \ell \nu_{\ell}}}
\newcommand{\BDstarstarlnu}{\ensuremath{B \to D^{**} \ell \nu_{\ell}}}
\newcommand{\Bzeroe}{\ensuremath{B^0 \to D^- e^+ \nu_{e}}}
\newcommand{\Bzeromu}{\ensuremath{B^0 \to D^- \mu^+ \nu_{\mu}}}
\newcommand{\Be}{\ensuremath{B^+ \to \bar{D}^0 e^+ \nu_{e}}}
\newcommand{\Bmu}{\ensuremath{B^+ \to \bar{D}^0 \mu^+ \nu_{\mu}}}
\newcommand{\Bzerol}{\ensuremath{B^0 \to D^- \ell^+ \nu_{\ell}}}
\newcommand{\Bl}{\ensuremath{B^+ \to \bar{D}^0 \ell^+ \nu_{\ell}}}
\newcommand{\etaEW}{\ensuremath{\eta_\mathrm{EW}}}
\newcommand{\Gone}{\ensuremath{\mathcal{G}(1)}}
\newcommand{\VGE}{\ensuremath{\etaEW\Gone\Vcb}}
\newcommand{\VE}{\ensuremath{\etaEW\Vcb}}

\newcommand{\Btag}{\ensuremath{B_{\mathrm{tag}}}}
\newcommand{\nbout}{\ensuremath{o_\mathrm{tag}}}
\newcommand{\lgnbout}{\ensuremath{\log_{10}(\nbout)}}
\newcommand{\Mbc}{\ensuremath{M_{\mathrm{bc}}}}
\newcommand{\Mmiss}{\ensuremath{M_{\mathrm{miss}}^2}}
\newcommand{\dGdw}{\ensuremath{\Delta \Gamma / \Delta w}}
\newcommand{\dGidw}{\ensuremath{\Delta \Gamma_i / \Delta w}}
\newcommand{\dGjdw}{\ensuremath{\Delta \Gamma_j / \Delta w}}
\newcommand{\dGiBGLdw}{\ensuremath{\Delta \Gamma_{i,\mathrm{BGL}} / \Delta w}}
\newcommand{\dGiCLNdw}{\ensuremath{\Delta \Gamma_{i,\mathrm{CLN}} / \Delta w}}

\newcommand{\dGidwf}{\ensuremath{\frac{\Delta \Gamma_i}{ \Delta w}}}
\newcommand{\dGjdwf}{\ensuremath{\frac{\Delta \Gamma_j}{ \Delta w}}}
\newcommand{\dGiCLNdwf}{\ensuremath{\frac{\Delta \Gamma_{i,\mathrm{CLN}}}{ \Delta w}}}
\newcommand{\dGiBGLdwf}{\ensuremath{\frac{\Delta \Gamma_{i,\mathrm{BGL}}}{ \Delta w}}}
\newcommand{\dGjBGLdwf}{\ensuremath{\frac{\Delta \Gamma_{j,\mathrm{BGL}}}{ \Delta w}}}
\newcommand{\dGjCLNdwf}{\ensuremath{\frac{\Delta \Gamma_{j,\mathrm{CLN}}}{ \Delta w}}}

\newcommand{\pics}{./fig}

\begin{document}

\preprint{\vbox{ \hbox{   }
                        \hbox{}
                        \hbox{}
			\hbox{Belle Preprint 2015-16}
                        \hbox{KEK Preprint 2015-43}
                        \hbox{}
}}

\quad\\[1.0cm]
\title{Measurement of the decay {\BDlnu} in fully reconstructed events and determination of the Cabibbo--Kobayashi--Maskawa matrix element {\Vcb}}

\noaffiliation
\affiliation{University of the Basque Country UPV/EHU, 48080 Bilbao}
\affiliation{University of Bonn, 53115 Bonn}
\affiliation{Budker Institute of Nuclear Physics SB RAS, Novosibirsk 630090}
\affiliation{Faculty of Mathematics and Physics, Charles University, 121 16 Prague}
\affiliation{Chonnam National University, Kwangju 660-701}
\affiliation{University of Cincinnati, Cincinnati, Ohio 45221}
\affiliation{Deutsches Elektronen--Synchrotron, 22607 Hamburg}
\affiliation{II. Physikalisches Institut, Georg-August-Universit\"at G\"ottingen, 37073 G\"ottingen}
\affiliation{SOKENDAI (The Graduate University for Advanced Studies), Hayama 240-0193}
\affiliation{Hanyang University, Seoul 133-791}
\affiliation{University of Hawaii, Honolulu, Hawaii 96822}
\affiliation{High Energy Accelerator Research Organization (KEK), Tsukuba 305-0801}
\affiliation{IKERBASQUE, Basque Foundation for Science, 48013 Bilbao}
\affiliation{Indian Institute of Technology Bhubaneswar, Satya Nagar 751007}
\affiliation{Indian Institute of Technology Guwahati, Assam 781039}
\affiliation{Indian Institute of Technology Madras, Chennai 600036}
\affiliation{Indiana University, Bloomington, Indiana 47408}
\affiliation{Institute of High Energy Physics, Chinese Academy of Sciences, Beijing 100049}
\affiliation{Institute of High Energy Physics, Vienna 1050}
\affiliation{Institute for High Energy Physics, Protvino 142281}
\affiliation{INFN - Sezione di Torino, 10125 Torino}
\affiliation{J. Stefan Institute, 1000 Ljubljana}
\affiliation{Kanagawa University, Yokohama 221-8686}
\affiliation{Institut f\"ur Experimentelle Kernphysik, Karlsruher Institut f\"ur Technologie, 76131 Karlsruhe}
\affiliation{Kennesaw State University, Kennesaw GA 30144}
\affiliation{King Abdulaziz City for Science and Technology, Riyadh 11442}
\affiliation{Department of Physics, Faculty of Science, King Abdulaziz University, Jeddah 21589}
\affiliation{Korea Institute of Science and Technology Information, Daejeon 305-806}
\affiliation{Korea University, Seoul 136-713}
\affiliation{Kyungpook National University, Daegu 702-701}
\affiliation{\'Ecole Polytechnique F\'ed\'erale de Lausanne (EPFL), Lausanne 1015}
\affiliation{Faculty of Mathematics and Physics, University of Ljubljana, 1000 Ljubljana}
\affiliation{Ludwig Maximilians University, 80539 Munich}
\affiliation{Luther College, Decorah, Iowa 52101}
\affiliation{University of Maribor, 2000 Maribor}
\affiliation{Max-Planck-Institut f\"ur Physik, 80805 M\"unchen}
\affiliation{School of Physics, University of Melbourne, Victoria 3010}
\affiliation{Moscow Physical Engineering Institute, Moscow 115409}
\affiliation{Moscow Institute of Physics and Technology, Moscow Region 141700}
\affiliation{Graduate School of Science, Nagoya University, Nagoya 464-8602}
\affiliation{Kobayashi-Maskawa Institute, Nagoya University, Nagoya 464-8602}
\affiliation{Nara Women's University, Nara 630-8506}
\affiliation{National Central University, Chung-li 32054}
\affiliation{National United University, Miao Li 36003}
\affiliation{Department of Physics, National Taiwan University, Taipei 10617}
\affiliation{H. Niewodniczanski Institute of Nuclear Physics, Krakow 31-342}
\affiliation{Nippon Dental University, Niigata 951-8580}
\affiliation{Niigata University, Niigata 950-2181}
\affiliation{Novosibirsk State University, Novosibirsk 630090}
\affiliation{Osaka City University, Osaka 558-8585}
\affiliation{Pacific Northwest National Laboratory, Richland, Washington 99352}
\affiliation{University of Pittsburgh, Pittsburgh, Pennsylvania 15260}
\affiliation{University of Science and Technology of China, Hefei 230026}
\affiliation{Soongsil University, Seoul 156-743}
\affiliation{Sungkyunkwan University, Suwon 440-746}
\affiliation{School of Physics, University of Sydney, NSW 2006}
\affiliation{Department of Physics, Faculty of Science, University of Tabuk, Tabuk 71451}
\affiliation{Tata Institute of Fundamental Research, Mumbai 400005}
\affiliation{Excellence Cluster Universe, Technische Universit\"at M\"unchen, 85748 Garching}
\affiliation{Department of Physics, Technische Universit\"at M\"unchen, 85748 Garching}
\affiliation{Toho University, Funabashi 274-8510}
\affiliation{Tohoku University, Sendai 980-8578}
\affiliation{Earthquake Research Institute, University of Tokyo, Tokyo 113-0032}
\affiliation{Department of Physics, University of Tokyo, Tokyo 113-0033}
\affiliation{Tokyo Institute of Technology, Tokyo 152-8550}
\affiliation{Tokyo Metropolitan University, Tokyo 192-0397}
\affiliation{University of Torino, 10124 Torino}
\affiliation{Utkal University, Bhubaneswar 751004}
\affiliation{CNP, Virginia Polytechnic Institute and State University, Blacksburg, Virginia 24061}
\affiliation{Wayne State University, Detroit, Michigan 48202}
\affiliation{Yamagata University, Yamagata 990-8560}
\affiliation{Yonsei University, Seoul 120-749}
  \author{R.~Glattauer}\affiliation{Institute of High Energy Physics, Vienna 1050} 
  \author{C.~Schwanda}\affiliation{Institute of High Energy Physics, Vienna 1050} 
  \author{A.~Abdesselam}\affiliation{Department of Physics, Faculty of Science, University of Tabuk, Tabuk 71451} 
  \author{I.~Adachi}\affiliation{High Energy Accelerator Research Organization (KEK), Tsukuba 305-0801}\affiliation{SOKENDAI (The Graduate University for Advanced Studies), Hayama 240-0193} 
  \author{K.~Adamczyk}\affiliation{H. Niewodniczanski Institute of Nuclear Physics, Krakow 31-342} 
  \author{H.~Aihara}\affiliation{Department of Physics, University of Tokyo, Tokyo 113-0033} 
  \author{S.~Al~Said}\affiliation{Department of Physics, Faculty of Science, University of Tabuk, Tabuk 71451}\affiliation{Department of Physics, Faculty of Science, King Abdulaziz University, Jeddah 21589} 
  \author{D.~M.~Asner}\affiliation{Pacific Northwest National Laboratory, Richland, Washington 99352} 
  \author{T.~Aushev}\affiliation{Moscow Institute of Physics and Technology, Moscow Region 141700} 
  \author{R.~Ayad}\affiliation{Department of Physics, Faculty of Science, University of Tabuk, Tabuk 71451} 
% ...
\collaboration{The Belle Collaboration}

\begin{abstract}
We present a determination of the magnitude of the Cabibbo--Kobayashi--Maskawa matrix element {\Vcb} using the decay {\BDlnu} 
($\ell=e,\mu$) based on 711{\invfb} of $e^+e^-\to \Upsilon(4S)$~data recorded by the Belle detector and containing $772 \times 10^6$ $B\bar{B}$ pairs. 
One $B$~meson in the event is fully reconstructed in a hadronic decay mode while the other, on the signal side, is partially reconstructed from a 
charged lepton and either a $D^+$ or $D^0$~meson in a total of 23 hadronic decay modes. 
The isospin-averaged branching fraction of the decay {\BDlnu} is found to be 
$\mathcal{B}(\Bzerol)=(2.31\pm 0.03(\mathrm{stat})\pm 0.11(\mathrm{syst}))\%$.
Analyzing the differential decay rate as a function of the hadronic recoil with the parameterization of Caprini, Lellouch and Neubert and using the 
form-factor prediction $\Gone=1.0541\pm 0.0083$ calculated by FNAL/MILC, we obtain $\VE=(40.12\pm 1.34)\times 10^{-3}$, where {\etaEW} is the 
electroweak correction factor. Alternatively, assuming the model-independent form-factor parameterization of Boyd, Grinstein and Lebed and using 
lattice QCD data from the FNAL/MILC and HPQCD collaborations, we find $\VE=(41.10 \pm 1.14)\times 10^{-3}$.
\end{abstract}

\maketitle

{\renewcommand{\thefootnote}{\fnsymbol{footnote}}}
\setcounter{footnote}{0}

\section{Introduction}
\label{sec:introduction}

The magnitude of the Cabibbo--Kobayashi--Maskawa~\cite{cabibbo1963unitary,Kobayashi:1973fv} matrix element {\Vcb} can be determined from 
inclusive semileptonic decays to charm final states {\BXclnu}~\cite{Gambino_2013rza} and from exclusive decays 
{\BDstarlnu} ~\cite{Dungel:2010uk,Aubert:2007rs} and {\BDlnu}~\cite{Aubert:2009ac}. Exclusive and inclusive measurements differ by 
about two to three standard deviations, where the current world averages determined by the Heavy Flavor Averaging Group~\cite{HFAG_averages} yield 
$\Vcb_{\BDstarlnu} = (38.94 \pm 0.76 ) \times 10^{-3}$ and $\Vcb_{\BXclnu} = (42.46 \pm 0.88) \times 10^{-3}$. 
The inclusive and exclusive (from {\BDstarlnu}) determinations of {\Vcb} are thus known with a precision of about 2\%.
Determinations of {\Vcb} with the decay {\BDlnu} are currently less precise with a world average of  
$\Vcb_{\BDlnu} = (39.45 \pm 1.67 ) \times 10^{-3}$; the 4\% error is dominated by the experimental uncertainty.
The main motivation of our study is to improve the determination of {\Vcb} from {\BDlnu} and thereby clarify the experimental 
knowledge of {\Vcb}.

The kinematics of the decay {\BDlnu} are described by the recoil variable $w$, defined as the product of the 4-velocities of
the $B$ and $D$ mesons. This quantity is related to the squared 4-momentum transfer to the lepton-neutrino system $q^2=(P_\ell+P_\nu)^2$:
\begin{equation}
  w=V_B\cdot V_D=\frac{m_B^2+m_D^2-q^2}{2m_B m_D}~,
\end{equation}
where $V_B$ and $V_D$ are the four-vector velocities of the $B$ and $D$ meson respectively, and $m_B$ and $m_D$ are their nominal masses 
\cite{footnote_c}. The minimum value of $w=1$ corresponds to zero recoil of the $D$~meson in the $B$ rest frame; the maximum value of $w$ corresponds 
to no 4-momentum transfer to the lepton-neutrino system ($q^2=0$):
\begin{equation}
  w_\mathrm{max} = \frac{m_B^2+m_D^2}{2m_Bm_D} \approx 1.6~.
\end{equation}
Using the latest measurements of $B$ and $D$ meson masses \cite{Agashe:2014kda}, this results in $w_\mathrm{max} (B^\pm) = 1.59209 \pm 0.00010$ for 
charged $B$ mesons and $w_\mathrm{max} (B^0) = 1.58901 \pm 0.00011$ for neutral $B$ mesons.

In the Heavy Quark Effective Theory (HQET) description of the {\BDlnu} decay rate, the leptonic and hadronic currents factorize up to a small 
electroweak correction~\cite{pbf}:
\begin{equation} \label{eq:vxb:dGamma}
d\Gamma \propto G_\mathrm{F}^2\lvert V_{cb}\rvert^2~\bigl\lvert L_\mu \langle
D \lvert \bar{c} \gamma^\mu b \rvert B \rangle \bigr\rvert^2 ~,
\end{equation}
where $G_\mathrm{F}$ is the Fermi coupling constant. The hadronic current is conventionally decomposed in terms of the vector and scalar form factors 
$f_+(q^2)$ and $f_0(q^2)$ as
\begin{equation}
\langle D \lvert \bar{c} \gamma^\mu b \rvert B \rangle = f_+(q^2) \left[ (P_B + P_D )^\mu - \frac{m_B^2 - m_D^2}{q^2} q^\mu \right]  
 + f_0(q^2) \frac{m_B^2 - m_D^2}{q^2} q^\mu.
\end{equation}
In the limit of negligible lepton masses, the differential decay rate does not depend on $f_0(q^2)$ and 
can be written as
\begin{equation}
  \frac{d\Gamma}{dw} = \frac{G_\mathrm{F}^2 m^3_{D}}{48\pi^3}(m_B+m_D)^2(w^2-1)^{3/2}
                     \eta_\mathrm{EW}^2{\Vcb}^2 |\mathcal{G}(w)|^2~, 
  \label{eq:dGammadw}
\end{equation}
in which the form factor $\mathcal{G}(w)$~\cite{neubert_1991} is given by
\begin{equation}
 \mathcal{G}(w)^2 = \frac{4r}{(1+r)^2} f_+(w)^2 ,
 \label{eq:fplusAndG}
\end{equation} 
where $r=m_D/m_B$ and $\eta_\mathrm{EW}$ is the electroweak correction that, at leading order, is 1.0066~\cite{Sirlin1982}.
While the measured decay rate depends only on $f_+$, theoretical calculations are also available for $f_0$ and can be included in the determination 
of {\Vcb} by using the kinematic constraint at maximum recoil $w_\mathrm{max}\approx 1.6$,
\begin{equation}
  f_0(w_\mathrm{max}) = f_+(w_\mathrm{max})~. \label{eq:kinematic}
\end{equation}

Different parameterizations of the form factor $\mathcal{G}(w)$ are available in the literature. 
A model-independent one that relies only on QCD dispersion relations has been proposed by Boyd, Grinstein and Lebed 
(BGL)~\cite{boyd1995constraints}:
\begin{equation}
 f_i(z) = \frac{1}{P_i(z) \phi_i(z)} \sum\limits_{n=0}^{N} a_{i,n} z^n, \quad i=+,0
 \label{eq:BGL}
\end{equation}
where
\begin{equation}
  z(w) = \frac{\sqrt{w + 1} - \sqrt{2}  }{\sqrt{w + 1} + \sqrt{2}  }~,
\end{equation}
$P_i(z)$ are the ``Blaschke factors'' containing explicit poles ({\it e.g.}, the $B_c$ or $B^*_c$ poles) in $q^2$ and $\phi_i(z)$ are the 
``outer functions,'' which are arbitrary but required to be analytic without any poles or branch cuts. The $a_{i,n}$ are free parameters and $N$ is 
the order at which the series is truncated. Following Ref.~\cite{lattice2015b}, we choose $P_i(z)=1$ and
\begin{eqnarray}
\phi_+(z) & = & 1.1213 (1+z)^2 (1-z)^{1/2}
           [(1+r)(1-z)+ 2\sqrt{r}(1+z)]^{-5}~, \label{eq:phiplus}\\
\phi_0(z) & = & 0.5299 (1+z)(1-z)^{3/2}  
	   [(1+r)(1-z)+ 2\sqrt{r}(1+z)]^{-4}~. \label{eq:phizero}
\end{eqnarray}
With this choice of the outer functions, the unitarity bound on the coefficients $a_{i,n}$ takes the simple form
\begin{equation}
  \sum\limits_{n=0}^{N} |a_{i,n}|^2 \leq 1~,
  \label{eq:unitarity}  
\end{equation}
for any order $N$.

The most commonly used form factor parametrization is the one of Caprini, Lellouch and Neubert (CLN)~\cite{Caprini_et_al}. 
It reduces the free parameters by adding multiple dispersive constraints and spin- and heavy-quark symmetries:
\begin{equation}
  \mathcal{G}(z)= \mathcal{G}(1)(1 - 8 \rho^2 z + (51 \rho^2 - 10 )
  z^2 - (252 \rho^2 - 84 ) z^3)~.
  \label{eq:CLN}
\end{equation}
The free parameters are the form factor at zero recoil $\mathcal{G}(1)$ and the linear slope $\rho^2$. The precision of this approximation 
is estimated to be better than 2\%, which is close to the current experimental accuracy of {\Vcb}.

The paper is organized as follows. In Sect.~\ref{sec:Experimental procedure}, we explain the details of our analysis procedure. 
In Sect.~\ref{sec:results}, we present our results and their systematic uncertainties.
Finally, in Sect.~\ref{sec:Discussion}, we interpret the differential {\BDlnu} decay rate, $\Delta\Gamma/\Delta w$, to extract a value of {\VE}.

\section{Experimental Procedure}
\label{sec:Experimental procedure}

\subsection{Data sample}
\label{subsec:Data Sample and Event Selection}

The analysis is based on the entire Belle $\Upsilon(4S)$ data sample of 711{\invfb}, which corresponds to 772 million $B\bar{B}$ events. The Belle 
detector, located at the KEKB asymmetric-energy $e^+e^-$ collider~\cite{kurokawa2003overview}, is a large-solid-angle magnetic spectrometer that
consists of a silicon vertex detector (SVD), a 50-layer central drift chamber (CDC), an array of aerogel threshold Cherenkov counters (ACC), a 
barrel-like arrangement of time-of-flight scintillation counters (TOF), and an electromagnetic calorimeter comprised of CsI(Tl) crystals (ECL) 
located inside a superconducting solenoid coil that provides a 1.5~T magnetic field. An iron flux-return located outside of the coil is instrumented 
to detect $K_L^0$ mesons and to identify muons (KLM). Electron candidates are identified using the ratio of the energy detected in the ECL to the 
track momentum, the ECL shower shape, position matching between track and ECL cluster, the energy loss in the CDC, and the response of the ACC. Muons 
are identified based on their penetration range and transverse scattering in the KLM detector. In the momentum region relevant to this analysis, 
charged leptons are identified with an efficiency of about 90\% and the probability to misidentify a pion as an electron (muon) is 0.25\% 
(1.4\%)~\cite{hanagaki2002electron,abashian2002muon}. Charged kaons and pions are identified by a combination of the energy loss in the CDC, the 
Cherenkov light in the ACC, and the time of flight in the TOF. Further details on the Belle detector and reconstruction 
procedures are given in Ref.~\cite{TheBelleDetector}.

In this analysis, we use a sample of generic simulated $B\bar B$ Monte Carlo (MC) events equivalent to about five times the Belle data, generated with 
{\tt EvtGen}~\cite{lange2001evtgen}. Full detector simulation based on {\tt GEANT3}~\cite{brun1984geant} is applied. Final-state radiation is 
simulated with the {\tt PHOTOS} package \cite{barberio1994photos}. The decay {\BDlnu} is simulated using the {\tt HQET2} model of {\tt EvtGen}, which 
is based on the CLN parameterization.

The main background to {\BDlnu} is the decay {\BDstarlnu}, which is also modeled using the CLN form-factor parameterization. 
Semileptonic decays involving orbitally-excited charmed mesons, {\BDstarstarlnu}, are simulated using the model of Leibovich-Ligeti-Stewart-Wise 
(LLSW)~\cite{leibovich1998semileptonic}. Charmless semileptonic decays are modeled by a mixture of known exclusive decays and an inclusive model for 
$b\to u$ semileptonic transitions. We adjust a number of parameters in the MC to match the most recent experimental values~\cite{Agashe:2014kda}. 
Corrected parameters include the $\Upsilon(4S)$ width into $B^+B^-$ and $B^0\bar B^0$, the branching fractions of the hadronic $D$~meson decay modes 
used in the signal reconstruction (see Sect.~\ref{subsec:Signal reconstruction}), the {\BDstarlnu} and {\BDstarstarlnu} branching fractions and 
form factors, and both the branching fractions of known exclusive charmless $B$~decays and the total inclusive {\BXulnu} rate.

Hadronic events are selected based on the charged track multiplicity and the visible energy in the calorimeter. This selection is described in detail 
in Ref.~\cite{abe2001measurement}. To suppress events from $e^+e^-\to q\bar q$~continuum, we require the ratio of the second to the zeroth Fox-Wolfram 
moment $R_2$ to be less than 0.4~\cite{Fox:1978vu}.

\subsection{Hadronic tagging and tag calibration}
\label{subsec:Hadronic Tagging}

The first step in the analysis is the reconstruction of the hadronic decay of one $B$~meson ({\Btag}) in the $\Upsilon(4S)$ event. The Belle algorithm 
for full hadronic reconstruction~\cite{feindt2011hierarchical} forms charged {\Btag} candidates from 17 final states \cite{footnote_cc}
($D^{*0} \pi^- $,
$D^{*0} \pi^- \pi^0$,
$D^{*0} \pi^- \pi^- \pi^+$,
$D^0 \pi^- $,
$D^0 \pi^- \pi^0$,
$D^0 \pi^- \pi^- \pi^+$,
$D^{*0} D_s^{*-} $,
$D^{*0} D_s^- $,
$D^0 D_s^{*-} $,
$D^0 D_s^- $,
$J/\psi K^- $,
$J/\psi K^- \pi^+\pi^- $,
$D^0 K^- $,
$D^+ \pi^- \pi^- $,
$D^{*0} \pi^- \pi^- \pi^+\pi^0$,
$J/\psi K^- \pi^0$, and
$J/\psi K_S^0 \pi^- $)
and neutral {\Btag}~candidates from 15 final states 
($D^{*+} \pi^-$,
$D^{*+} \pi^- \pi^0$,
$D^{*+} \pi^- \pi^+ \pi^- $,
$D^+ \pi^- $,
$D^+ \pi^- \pi^0$,
$D^+ \pi^- \pi^+ \pi^- $,
$D^{*+} D_s^{*-} $,
$D^{*+} D_s^- $,
$D^+ D_s^{*-} $,
$D^+ D_s^- $,
$J/\psi K_S^0 $,
$J/\psi K^- \pi^+$,
$J/\psi K_S^0 \pi^+ \pi^- $,
$D^0 \pi^0 $, and
$D^{*+} \pi^- \pi^- \pi^+ \pi^0$).
To reconstruct the above $B$ decays, along with the subsequent hadronic decays of $D^{*0}$, $D^{*+}$, $D^0$, $D^+$, $D^{*+}_s$ and $D^+_s$
and lepton-pair decays of the $J/\psi$, the algorithm investigates 1104 different decay topologies. 
The selection of each decay chain is optimized using the neural network framework {\tt NeuroBayes}~\cite{NeuroBayes} and results in a multivariate 
classifier {\nbout}. Values of {\nbout} range from 0 to 1, where zero corresponds to background-like events and unity to signal-like events. Only 
candidates with $\nbout >10^{-3}$ are retained for further analysis. In addition to the selections already applied in the Belle full-reconstruction 
algorithm, we require the beam-energy constrained mass {\Mbc} of the {\Btag} candidate to be greater than 5.24~GeV, where {\Mbc} is defined as 
$\Mbc \equiv \sqrt{{E^2_\mathrm{beam}} -\Vec{p}_B^{\,2}}$.
Here, $E_\mathrm{beam}$ and $\vec p_B$ are the beam energy and the 3-momentum of the $B$~candidate in the $\Upsilon(4S)$ frame. If the signal $B$ 
candidate, described in the next section, is charged (neutral), we retain only charged (neutral) {\Btag} candidates. If an event has more than one 
possible {\Btag} candidate, we retain the candidate with the highest value of {\nbout}.

The Belle full-reconstruction tag requires calibration of its efficiency with data. Since the default Belle tag calibration \cite{BXulnu} uses 
{\BDlnu} decays, it can not be used in this analysis. We therefore derive an independent calibration based on fully reconstructed events in which the 
other $B$~meson decays semileptonically ({\BXlnu}). In addition to the selections already applied on {\Btag}, we require an identified lepton ($e$ or 
$\mu$) amongst the particles not used in the {\Btag} reconstruction.  The impact parameter relative to the $e^+e^-$ interaction point of the lepton in 
the plane perpendicular to the beam (along the beam) must be less than 0.5~cm (2~cm). The electron (muon) momentum in the laboratory frame is required 
to be greater than 0.3 GeV (0.6 GeV) and the polar angle relative to the beam axis of the lepton momentum in the same frame must lie in the range 
17-150$^\circ$ (25-145$^\circ$).
In electron events, we attempt to recover QED bremsstrahlung by searching for a photon within a 5$^\circ$ cone around the lepton direction. If such a 
photon is found, it is merged with the electron. If more than one photon satisfies this criterion, the photon closest to the lepton direction is 
chosen.

Separate calibration coefficients are derived for the 17 charged and 15 neutral {\Btag} modes. We further divide each calibration sample into 15 
equidistant bins in {\lgnbout} in the region between $-3$ and 0. In each calibration sample and in each {\lgnbout} bin, we count the number of 
events in the data and in the MC simulation (after scaling to the data luminosity and applying all corrections mentioned in 
Sect.~\ref{subsec:Data Sample and Event Selection}). We use the ratio of these yields as the calibration factor of the particular {\Btag}~mode in 
the {\lgnbout} bin. In total, 480 calibration coefficients are derived in this way. Overall, the calibration factor is around 0.8, with 90\% of the 
calibration factors lying between 0.5 and 1.1.

\subsection{Signal reconstruction}
\label{subsec:Signal reconstruction}

The {\BDlnu}~signal is reconstructed from the particles remaining in the event after excluding the charged tracks and photon candidates used in the 
reconstruction of {\Btag}. We require charged particles to have an impact parameter with respect to the interaction point of less than 0.5~cm (2~cm) 
in the plane perpendicular to the beam (along the beam), except for pions from $K^0_S \to \pi^+ \pi^-$~decays. Photon candidates in the event must 
have an energy greater than 50~MeV in the barrel region ($32^\circ<\theta<130^\circ$). In the forward (backward) endcap defined by 
$17^\circ<\theta<32^\circ$ ($130^\circ<\theta<150^\circ$), we require $E_\gamma>100$~(150)~MeV.

Amongst the particles remaining in the event, we search for identified electrons or muons for which we apply the momentum and polar-angle 
requirements described in Sect.~\ref{subsec:Hadronic Tagging}. We also recover QED bremsstrahlung by the algorithm described earlier.

Excluding the {\Btag} particles and the charged lepton, we search among the remaining particles in the event for $D^+$ decays into 10 final states 
($K^- \pi^+ \pi^+$,
$K^- \pi^+ \pi^+ \pi^0$,
$K^0_S \pi^+ $,
$K^0_S \pi^+ \pi^0 $,
$K^+ K^- \pi^+$,  
$K^0_S K^+ $,       
$K^0_S \pi^+ \pi^+ \pi^- $,
$\pi^+ \pi^0 $,
$\pi^+ \pi^+ \pi^- $, and
$K^- \pi^+ \pi^+ \pi^+ \pi^- $)
and $D^0$ decays into 13 final states 
($K^- \pi^+ $,
$K^- \pi^+ \pi^0 $, 
$K^- \pi^+ \pi^+ \pi^- $,
$K^0_S \pi^+ \pi^- $,
$K^0_S \pi^+ \pi^- \pi^0 $,
$K^0_S \pi^0 $,
$K^+ K^- $,
$\pi^+ \pi^- $,
$K^0_S K^0_S$,
$\pi^0 \pi^0$,
$K^0_S \pi^0 \pi^0$,
$K^- \pi^+ \pi^+ \pi^- \pi^0$, and
$\pi^+ \pi^- \pi^0$).
The branching fractions of the charged and neutral $D$ decay modes comprise 28.9\% and 40.1\% of the total rate, respectively~\cite{Agashe:2014kda}.

Neutral pions are reconstructed from photon pairs. We require the invariant mass of the two photons to lie within 15~MeV of the nominal $\pi^0$~mass 
(about 2.5 times the experimental resolution). All $\pi^0$~candidates satisfying this criterion are sorted according to the energy of their most 
energetic $\gamma$. If two pions share the most energetic $\gamma$, they are sorted by the energy of the second $\gamma$ in the pair. Starting from 
the most energetic combination, a $\pi^0$ candidate is removed if either of its photons has been used in a higher-ranked pion. We further require the 
opening angle of the two photons to be below 60$^{\circ}$ in the $e^+e^-$ center-of-mass frame.

$K^0_S$~mesons are reconstructed from their decay to two charged pions. We require the invariant two-pion mass to lie in the range 0.482-0.514~GeV
(a window of about 4 times the experimental resolution around the nominal mass). Different selections are applied depending on the momentum of the 
$K^0_S$~candidate in the laboratory frame: For low ($p<0.5$~GeV), medium ($0.5\leq p\leq 1.5$~GeV) and high momentum ($p>1.5$~GeV) candidates, we 
require the impact parameters of the pion daughters in the plane perpendicular to the beam to be greater than 0.05~cm, 0.03~cm, and 0.02~cm, 
respectively. The angle in the plane perpendicular to the beam between the vector from the interaction point to the $K^0_S$~vertex and the $K^0_S$ 
flight direction is required to be less than 0.3~rad, 0.1~rad and 0.03~rad for low, medium, and high momentum candidates, respectively; the separation 
distance of the two pion trajectories in the direction of the beam at their intersection point must be below 0.8~cm, 1.8~cm, and 2.4~cm, respectively. 
Finally, for medium (high) momentum $K^0_S$ candidates, we require the flight length in the plane perpendicular to the beam to be greater than 
0.08~cm (0.22~cm).

The invariant mass of a $D$~candidate is required to lie within $\pm 3$~standard deviations of the nominal $D^0$ or $D^+$~mass. We determine the width 
of the signal peak by fitting the reconstructed $D$~mass distribution separately in each channel.

We further reduce combinatorial background by requiring no unused charged particles in the event. The total energy in the event remaining in the ECL 
after excluding {\Btag}, the charged lepton, and the $D$~candidate must be below 1~GeV. The probability to reconstruct multiple combinations of 
{\Btag}, identified lepton, and $D$~candidate in the same event is very low ($<2\%$) so we do not apply a best-candidate selection in this analysis.

\subsection{Signal yield extraction}
\label{subsec:Yield extraction}

For the remainder of the analysis, we split the sample of selected events according to the lepton type and the charge of the {\Btag} candidate. 
Hereinafter, we refer to these sub-samples as {\Bzeroe}, {\Bzeromu}, {\Be}, and {\Bmu}.

In each sub-sample, we extract the {\BDlnu} signal yield from the distribution of the missing mass squared,
\begin{equation}
  \Mmiss=(P_\mathrm{LER}+P_\mathrm{HER}-P_{B_\mathrm{tag}}-P_D-P_\ell)^2~,
\end{equation}
where $P_\mathrm{HER}$ and $P_\mathrm{LER}$ are the 4-momenta of the colliding beams and $P_{B_\mathrm{tag}}$, $P_D$ and $P_\ell$ are the 4-momenta of
the \Btag, $D$, and charged-lepton candidates, respectively. For signal, the only missing particle is the neutrino of the {\BDlnu} decay and 
the missing-mass-squared distribution thus exhibits a prominent peak at zero. We determine the yield of this component by using a fit that accounts 
for the following contributions to the observed {\Mmiss}~distribution:
\begin{itemize}
\item {\it {\BDlnu} signal}: Events that contain a {\BDlnu} signal decay,
\item {\it {\BDstarlnu} background}: Events that contain a semileptonic $B$-meson decay to either a $D^{*+}$ or a $D^{*0}$ meson,
\item {\it Other backgrounds}: All events that do not fall in the aforementioned categories.
\end{itemize}

The resolution of the {\Mmiss} signal peak in real data is slightly worse than predicted by the MC simulation. We therefore add an additional 
Gaussian smearing of $(30 \pm 3.6)$~MeV$^2$ to the signal component in the MC, determined by comparing the signal peak width in data and MC.

\begin{figure*}
	\includegraphics[width=0.28\textwidth]{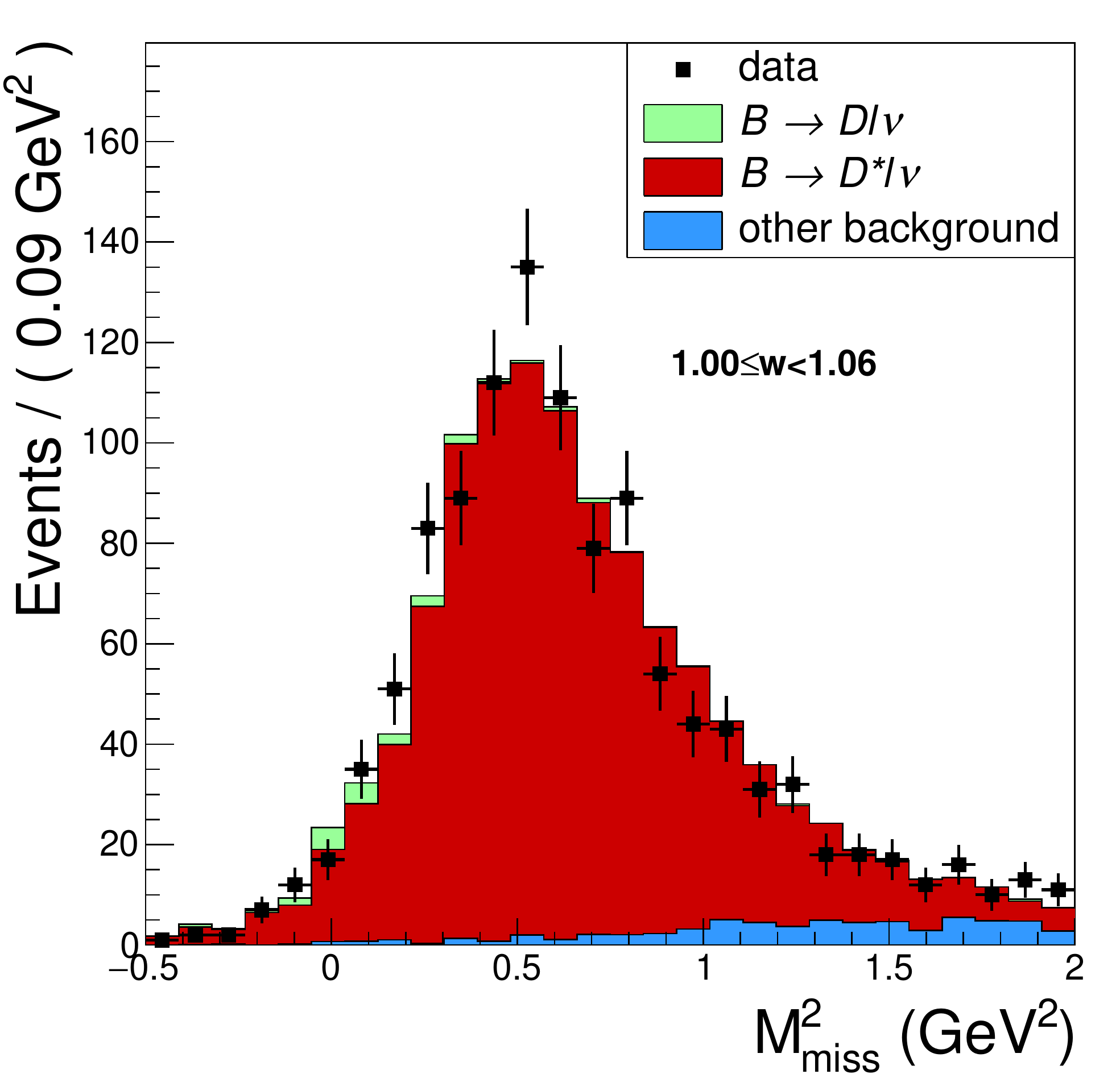}
	\includegraphics[width=0.28\textwidth]{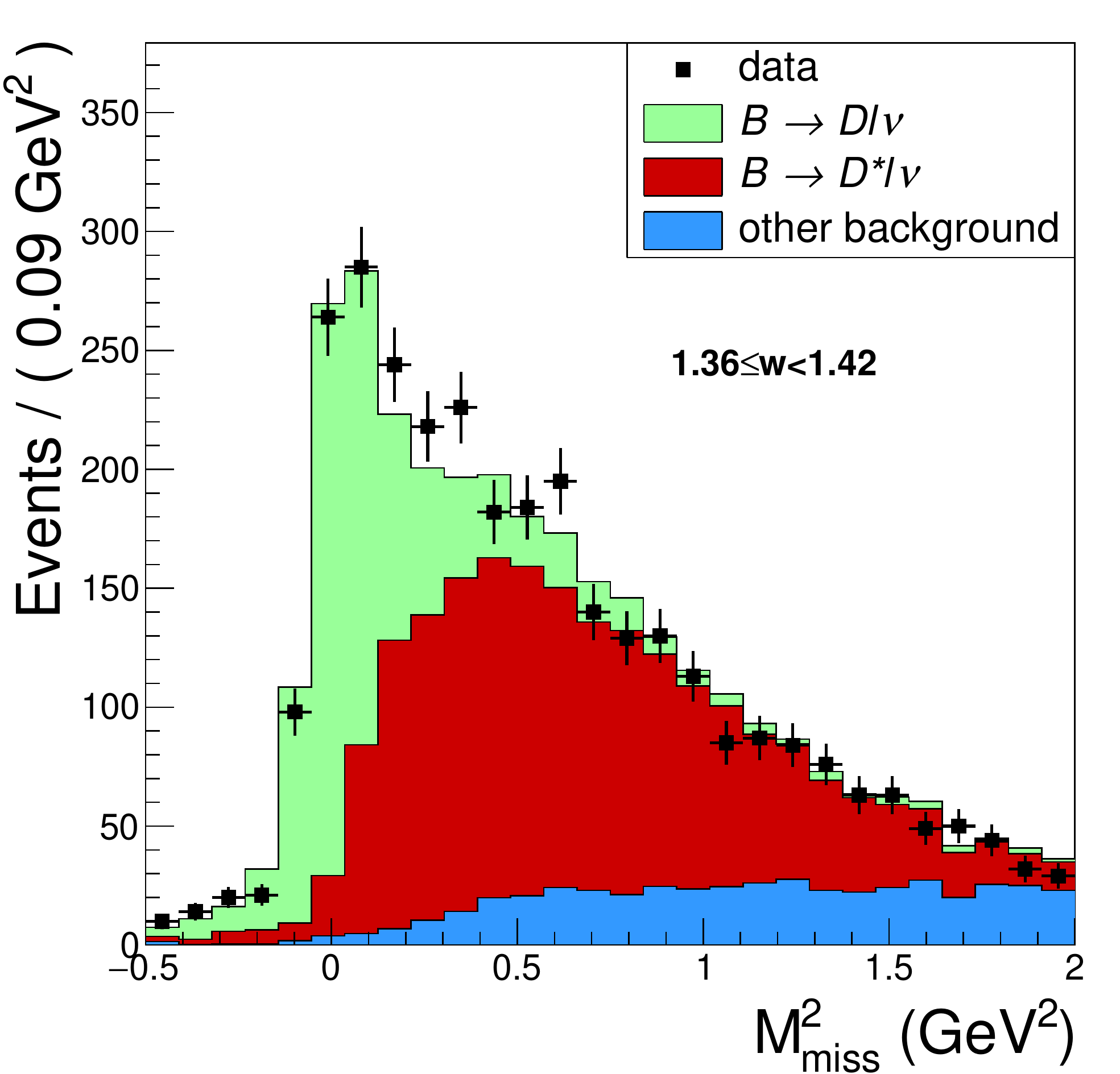}
	\includegraphics[width=0.28\textwidth]{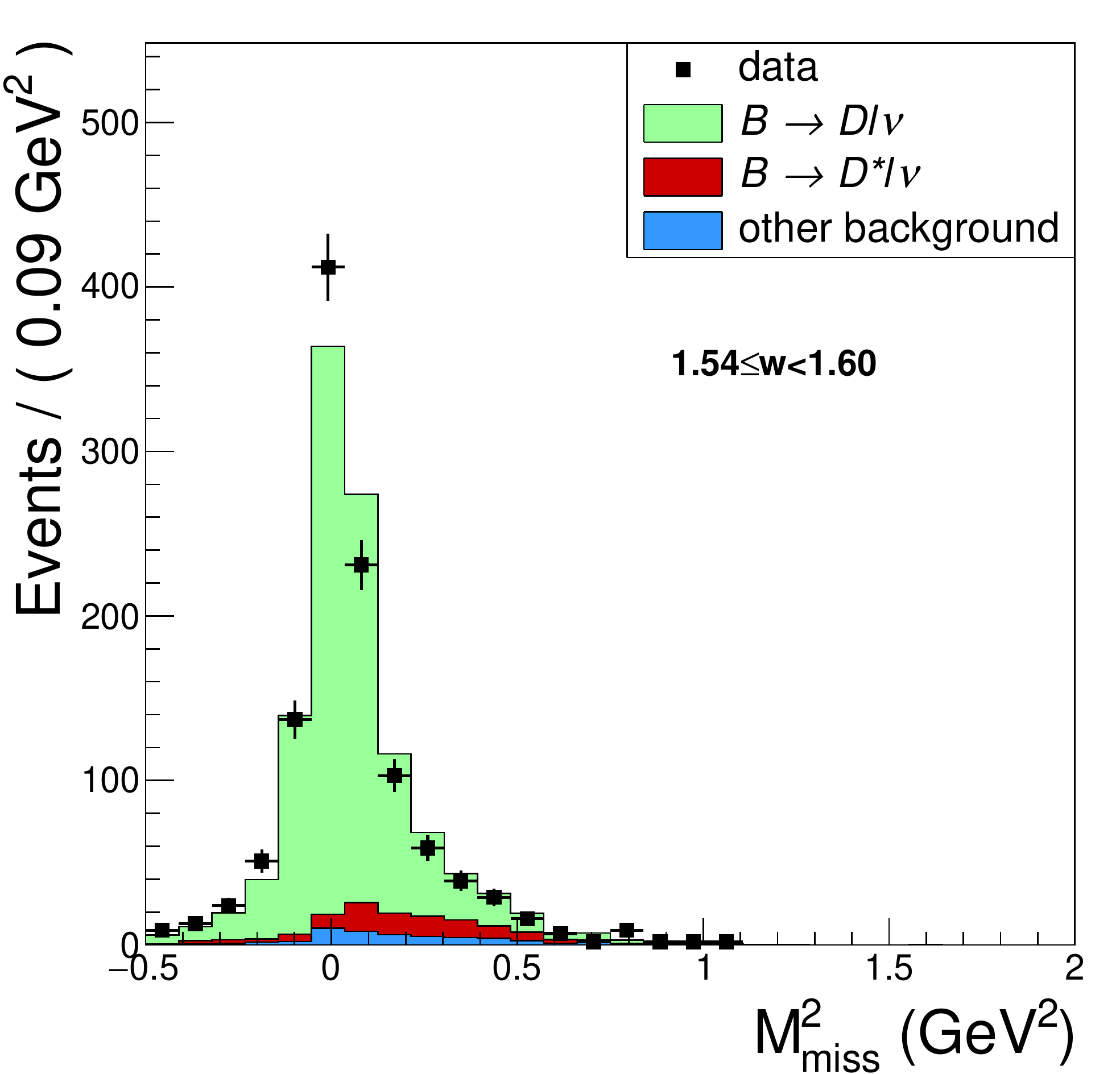}
\caption{(Color online) Fit to the missing mass squared distribution in three bins of $w$ for the {\Be} sub-sample. Points with error bars are the 
data. Histograms are (from top to bottom) the {\BDlnu} signal (green), the {\BDstarlnu} cross-feed background (red), and other backgrounds (blue). The 
$p$-values of the fits are (from left to right) 0.55, 0.21, and 0.10.} 
\label{fig:yield_fitBe}
	\includegraphics[width=0.28\textwidth]{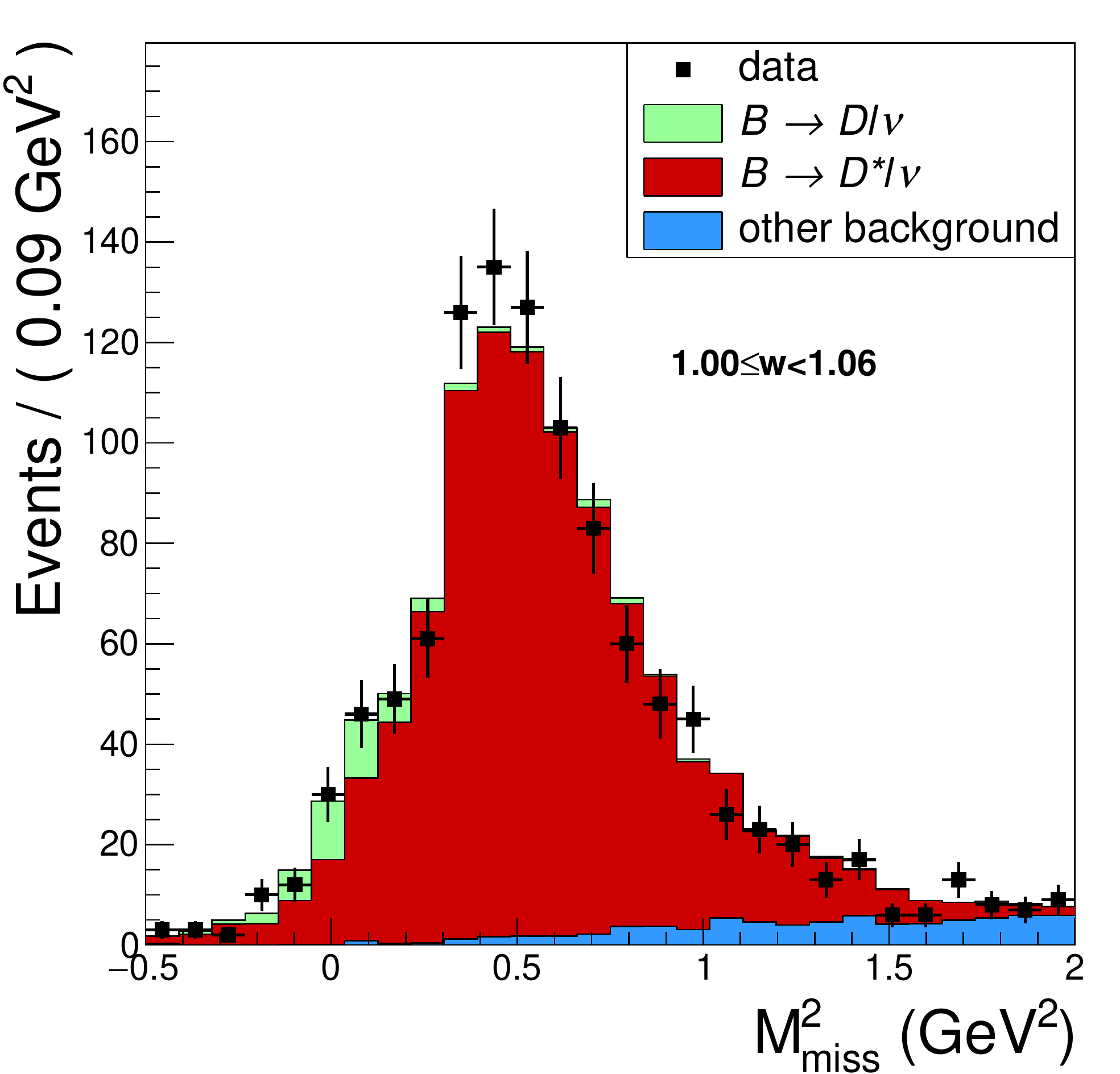}
	\includegraphics[width=0.28\textwidth]{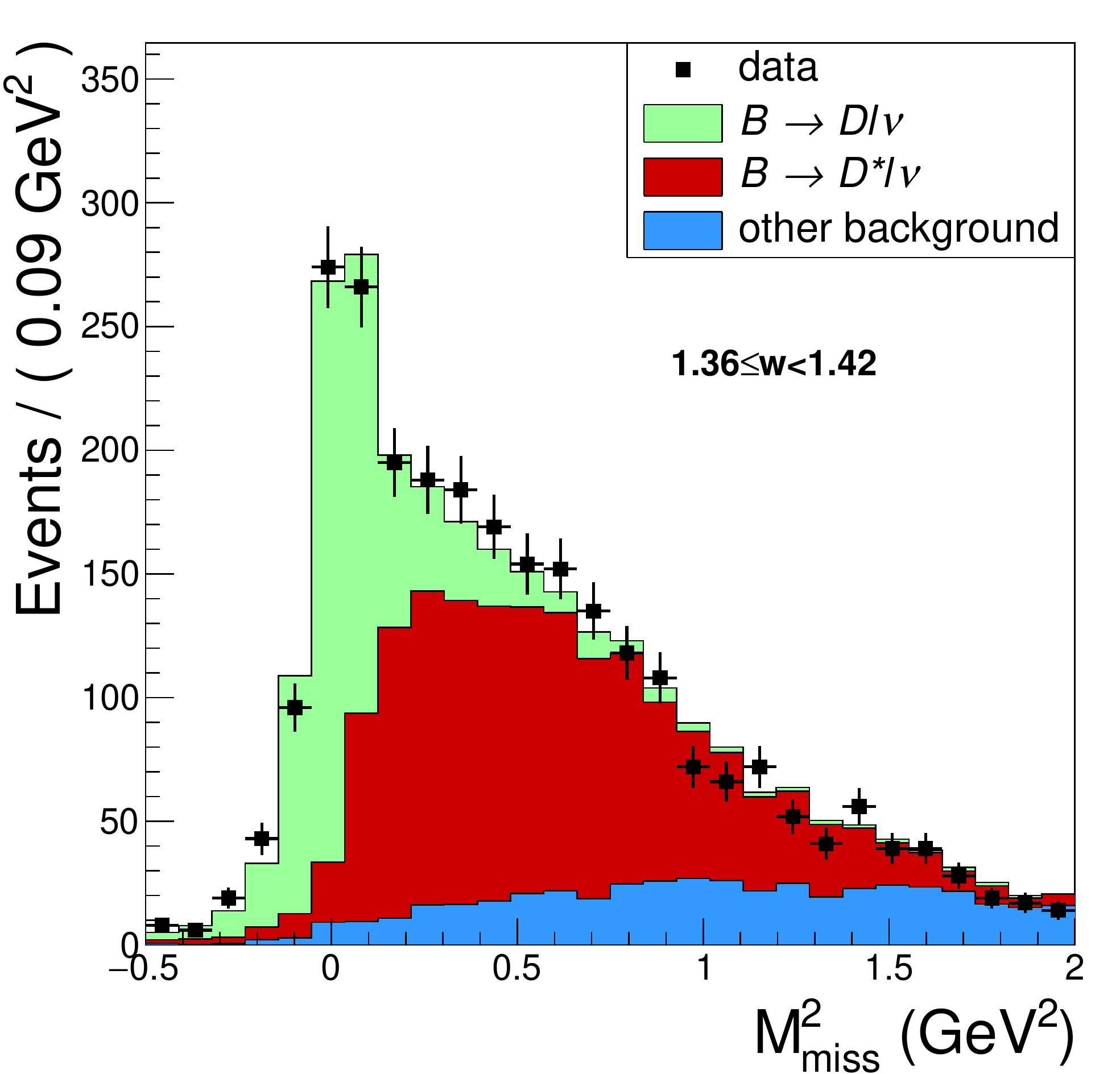}
	\includegraphics[width=0.28\textwidth]{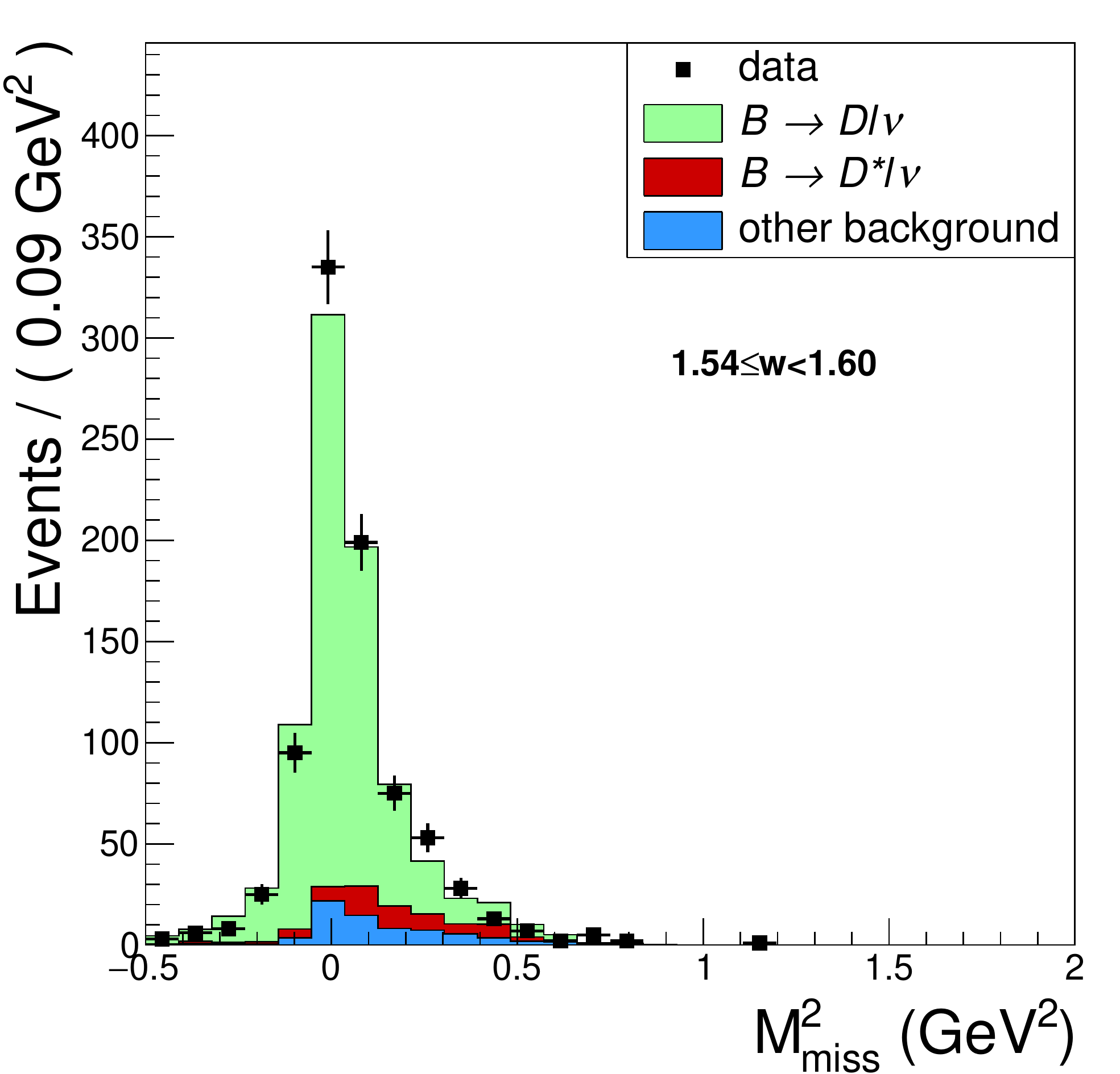}
\caption{Same as Fig.~\ref{fig:yield_fitBe} for the {\Bmu} sub-sample. The $p$-values of the fits are (from left to right) 0.71, 0.38, 
and 0.42.}
\label{fig:yield_fitBmu}
	\includegraphics[width=0.28\textwidth]{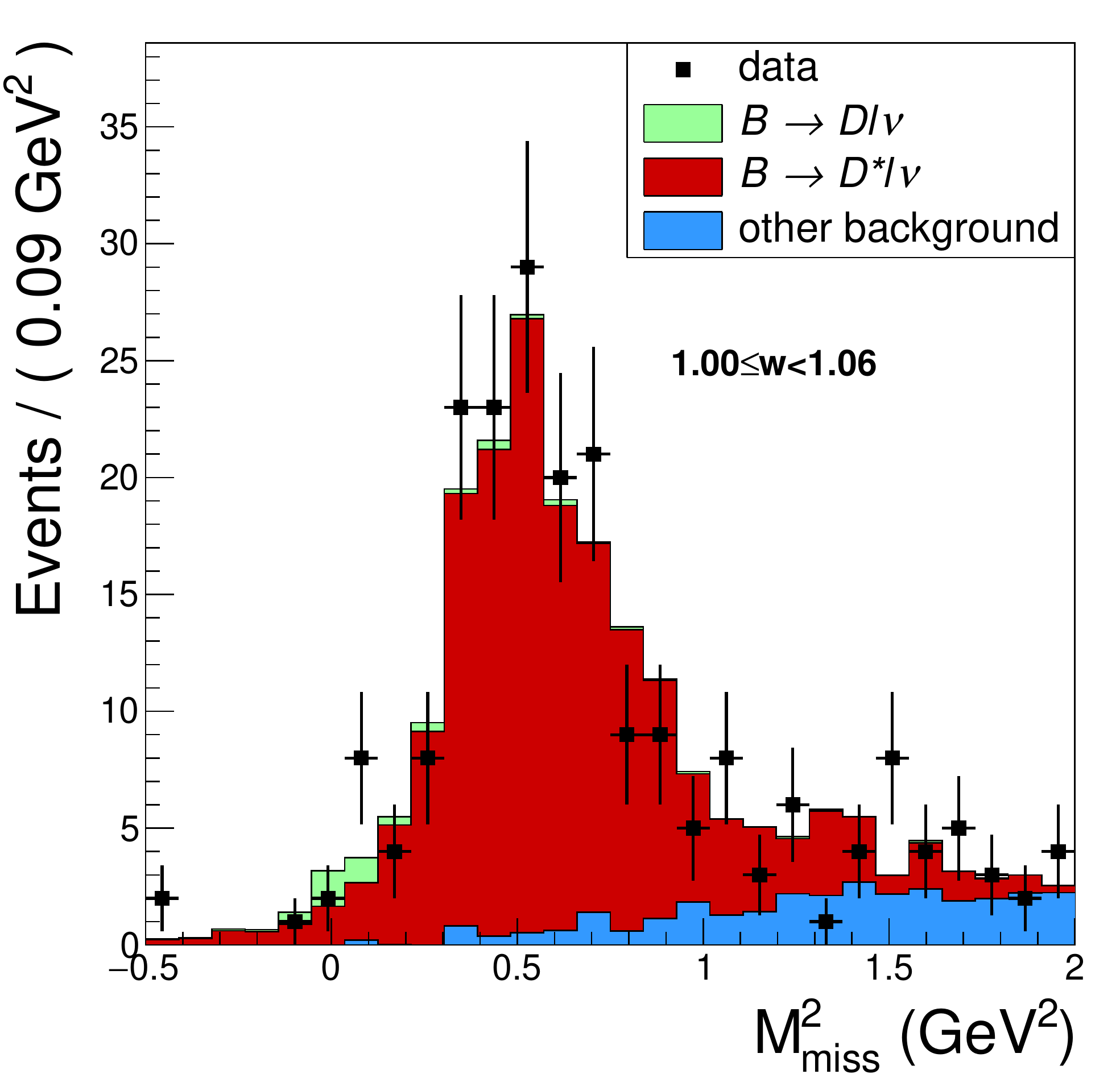}
	\includegraphics[width=0.28\textwidth]{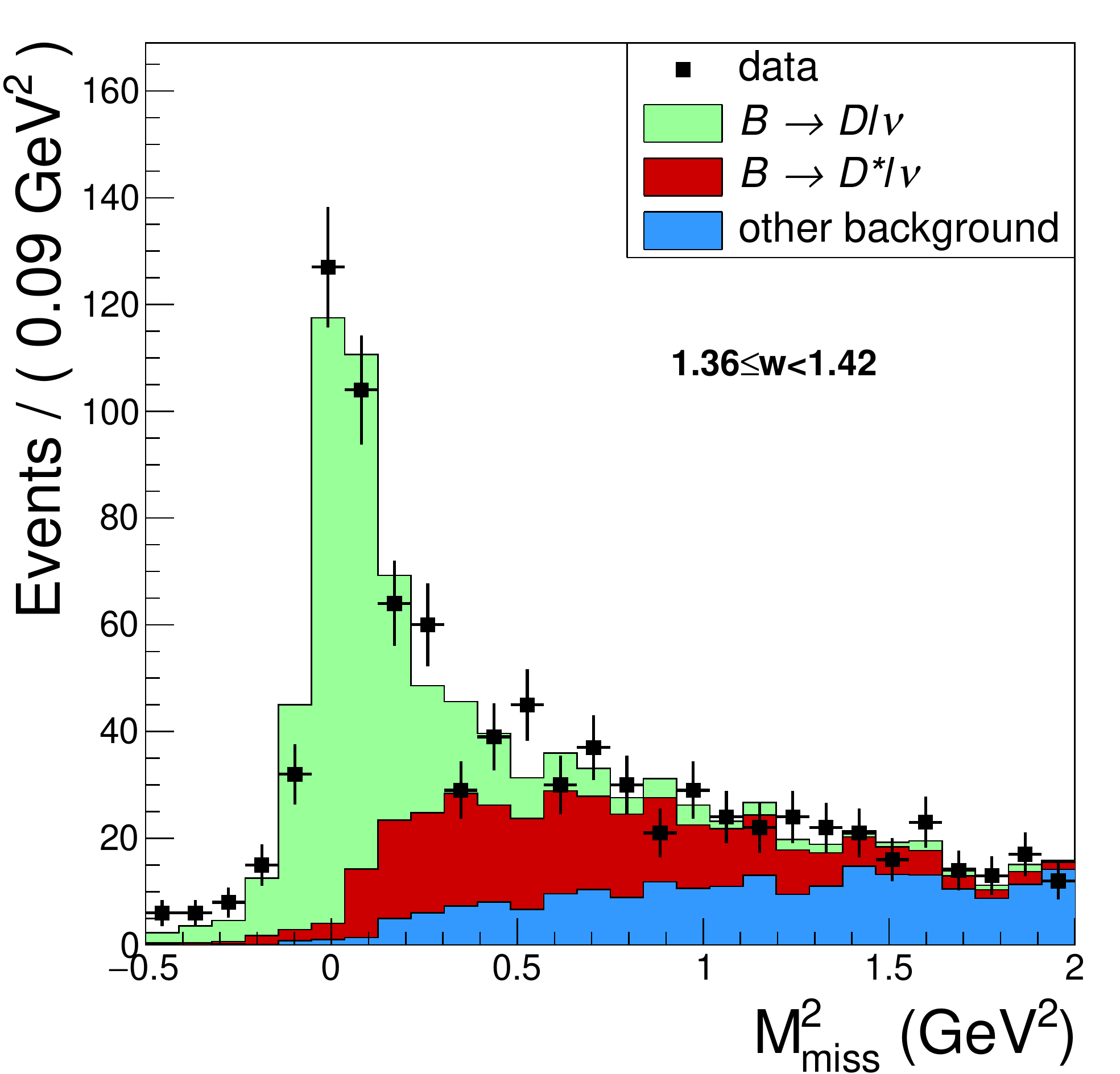}
	\includegraphics[width=0.28\textwidth]{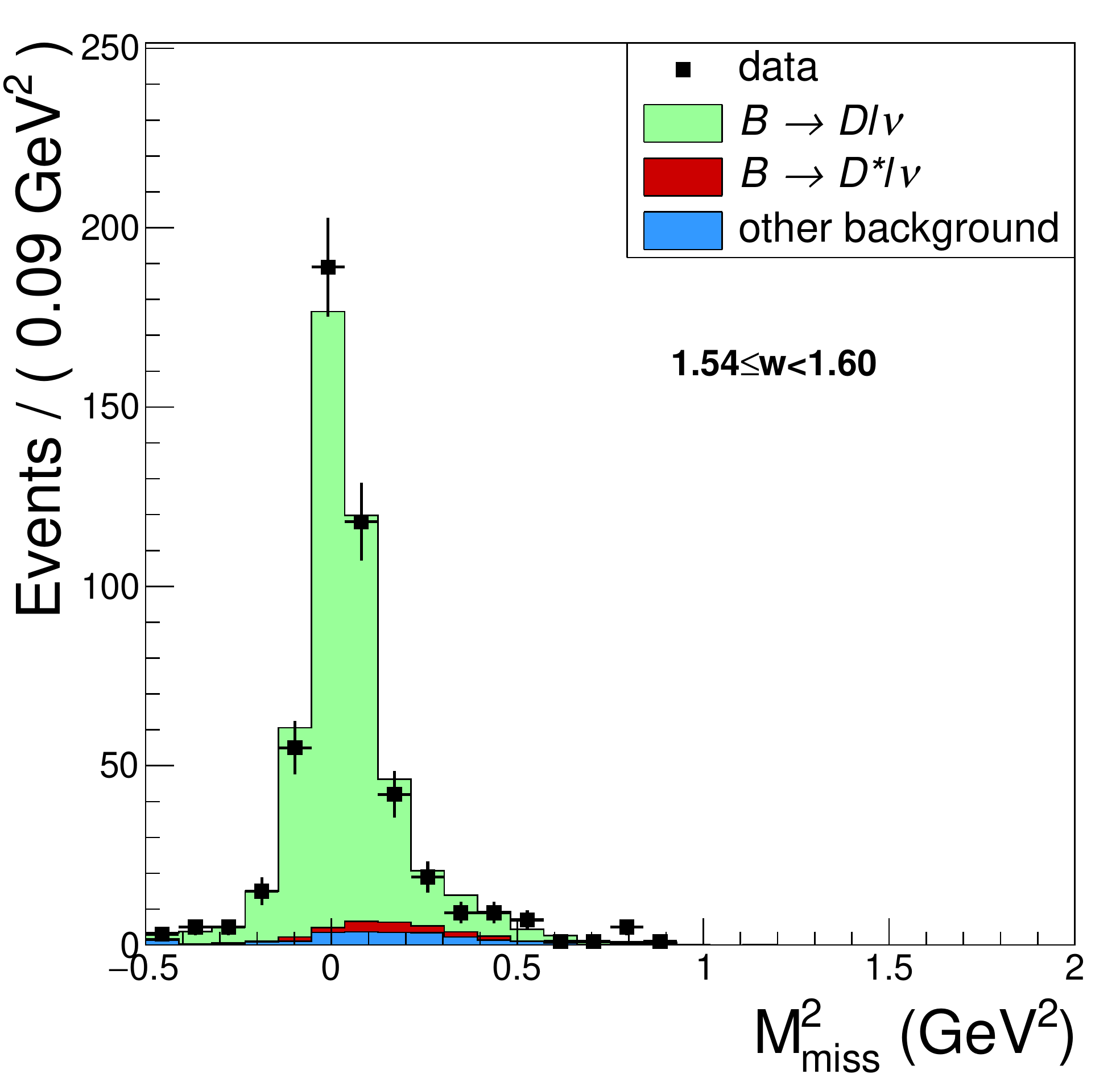}
\caption{Same as Fig.~\ref{fig:yield_fitBe} for the {\Bzeroe} sub-sample. The $p$-values of the fits are (from left to right) 0.30, 
0.10, and 0.96.}
\label{fig:yield_fitB0e}
	\includegraphics[width=0.28\textwidth]{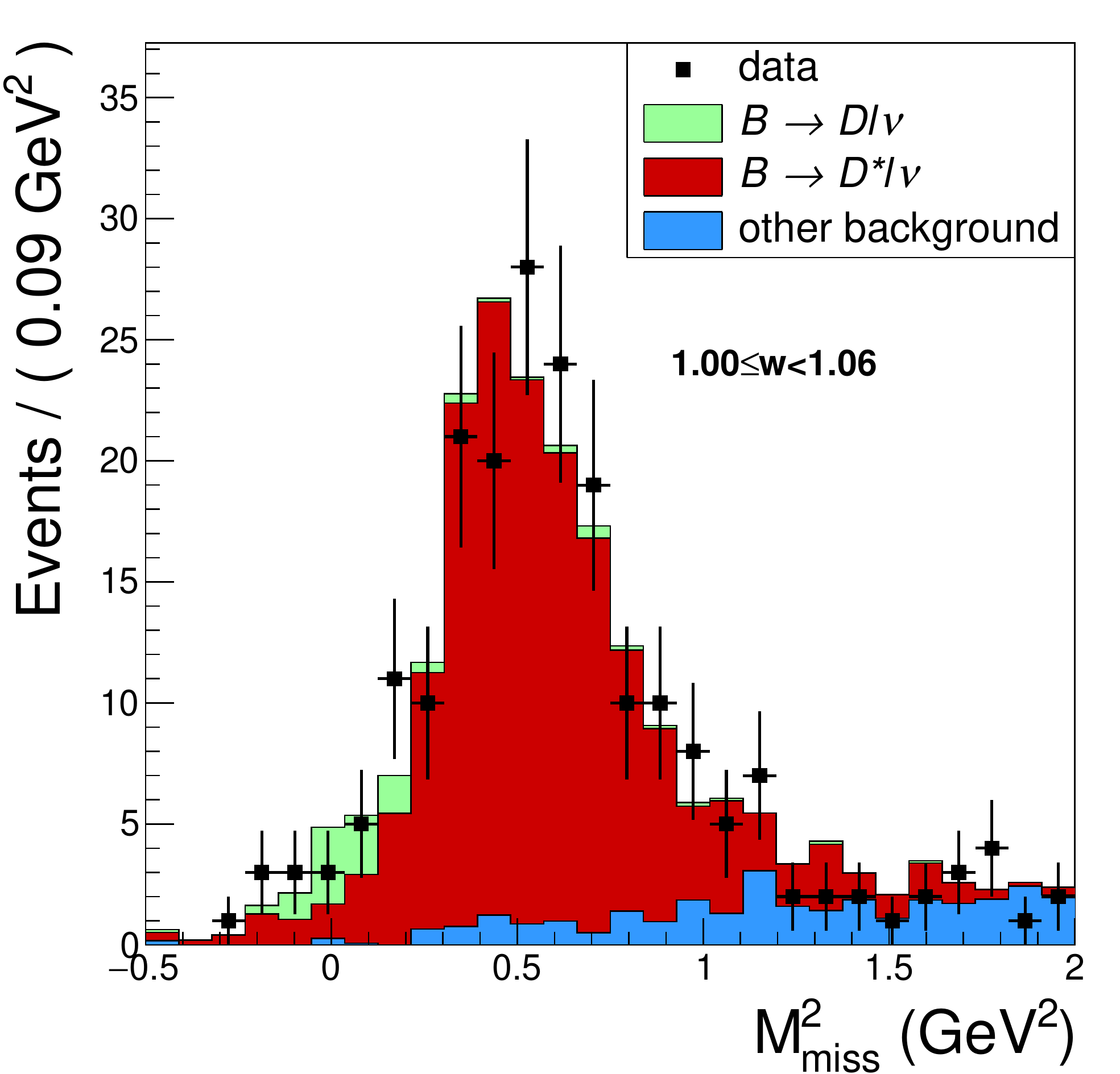}
	\includegraphics[width=0.28\textwidth]{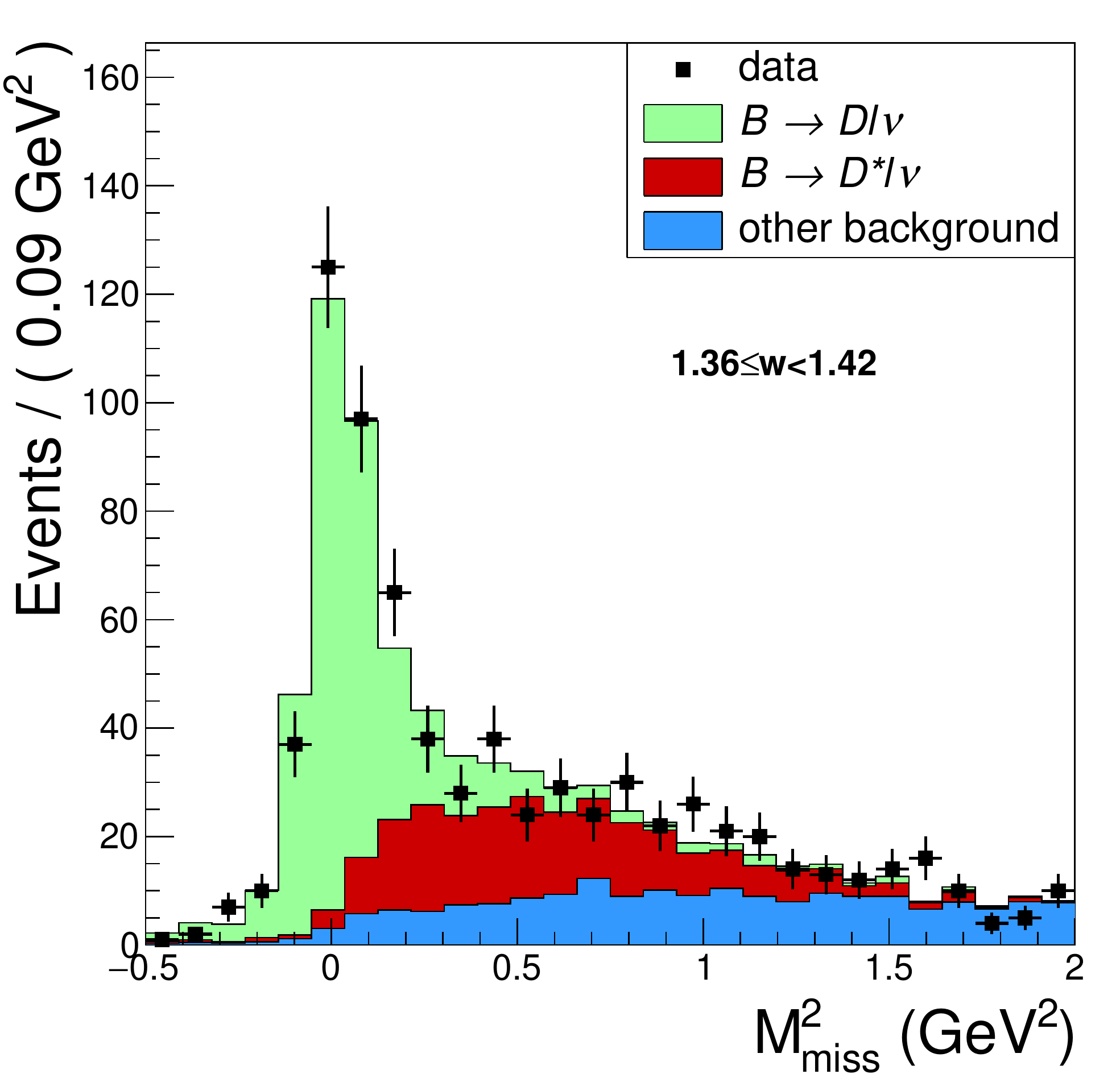}
	\includegraphics[width=0.28\textwidth]{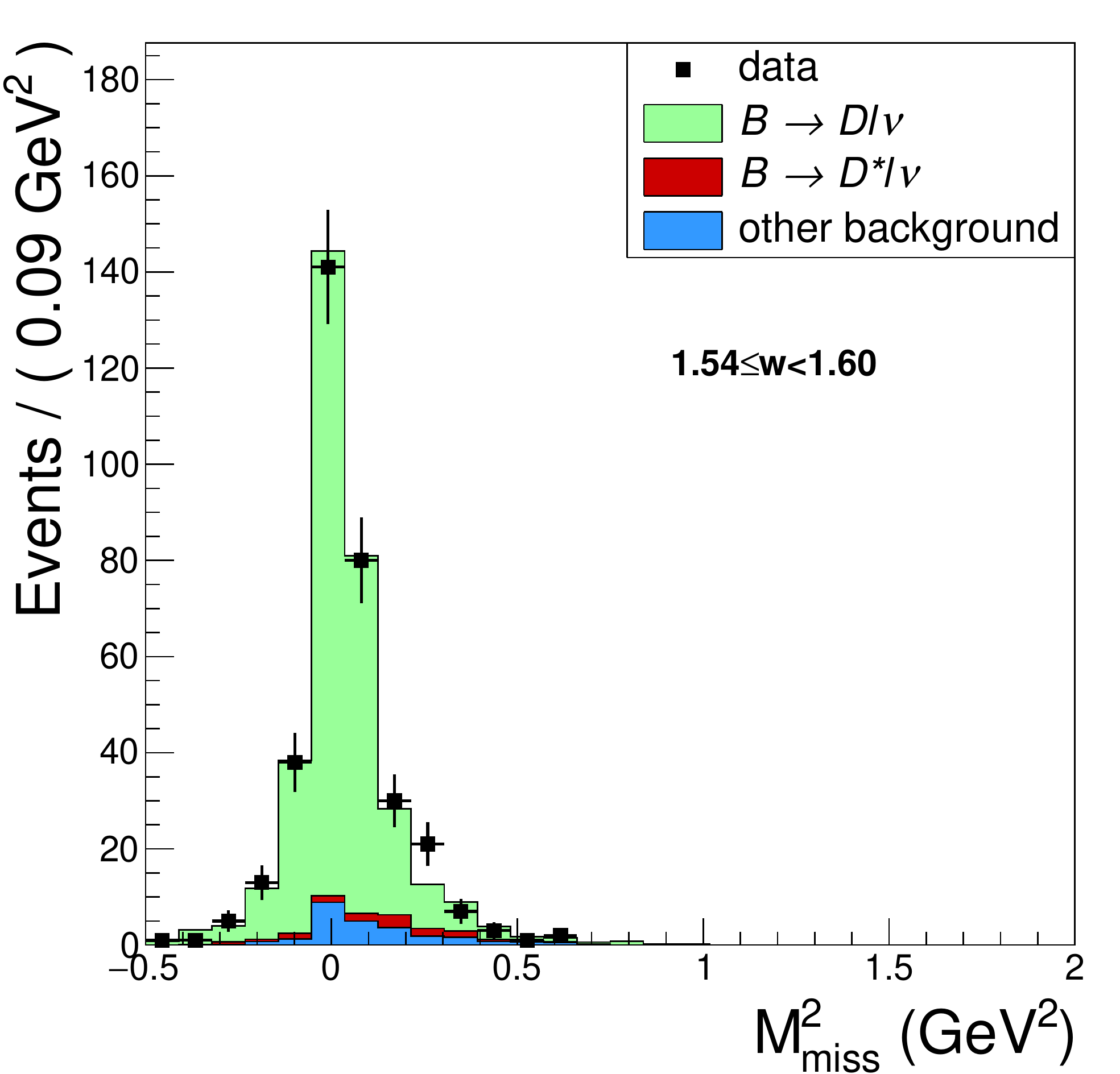}
\caption{Same as Fig.~\ref{fig:yield_fitBe} for the {\Bzeromu} sub-sample. The $p$-values of the fits are (from left to right) 0.92, 
0.39, and 1.00.}
\label{fig:yield_fitB0mu}
\end{figure*}

The fit uses the binned extended maximum likelihood algorithm by Barlow and Beeston~\cite{Fraction_Fit_Barlow} with MC templates obtained from 
simulation and takes into account the uncertainties of both data and MC templates.
This fit is performed separately in ten equal-size bins of $w$ in the range from 1 to 1.6. The bin width of $\Delta w=0.06$ is about an order of 
magnitude larger than the resolution in $w$ of about 0.005. Note that the kinematic endpoint of the $w$~distribution is slightly below the upper 
boundary of the last bin; the yield in the last bin drops for this reason. In every $w$~bin, the {\BDlnu} and {\BDstarlnu} components are allowed to 
float, while the other background component is small and is fixed to the MC expectation. Only in the last bin ($1.54<w<1.6$) is the {\BDstarlnu} 
component also fixed. The results of the fit in selected bins of $w$ are shown in Figs.~\ref{fig:yield_fitBe}~to~\ref{fig:yield_fitB0mu} for the 
{\Be}, {\Bmu}, {\Bzeroe}, and {\Bzeromu} sub-samples, respectively.

\section{Results and Systematic Uncertainties}
\label{sec:results}

\subsection{Results}

In each of the 4 sub-samples, we determine the differential decay width as a function of $w$ using
\begin{equation}
 \frac{\Delta\Gamma_i}{\Delta w}=\frac{\Delta\Gamma_{i,\mathrm{MC}}}{\Delta w}\frac{\tau_\mathrm{MC}}{\tau}\frac{N_i}{N_{i,\mathrm{MC}}}~, 
 \quad i=0,\dots,9 \,.
\label{eq:dGdw}
\end{equation}
Here, $\Delta\Gamma_{i,\mathrm{MC}}/\Delta w$ is the differential {\BDlnu} width expected in the $i^\mathrm{th}$ bin of $w$ assuming the values of 
the CLN parameters used in the MC:
\begin{equation}
 \frac{\Delta\Gamma_{i,\mathrm{MC}}}{\Delta w}=\frac{1}{\Delta w}\int_{w_{i,\mathrm{min}}}^{w_{i,\mathrm{max}}}\frac{d\Gamma_\mathrm{CLN}}{dw} dw~,
\end{equation}
where $w_{i,\mathrm{min}}$ and $w_{i,\mathrm{max}}$ are the boundaries of the $i^\mathrm{th}$~bin. Depending on the sub-sample, $\tau$ is the $B^+$ 
or $B^0$ lifetime ($\tau_{B^0}=1.519$~ps and $\tau_{B^+}=1.638$~ps, respectively~\cite{Agashe:2014kda}) and $\tau_\mathrm{MC}$ is the corresponding 
quantity in the MC simulation. Finally, $N_i$ is the {\BDlnu} signal yield measured by the missing-mass-squared fit in the $i^\mathrm{th}$ bin of $w$, 
and $N_{i,\mathrm{MC}}$ is the same quantity in the MC simulation after scaling to the data luminosity and applying all corrections mentioned in 
Sect.~\ref{subsec:Data Sample and Event Selection}. 

The results of {\dGidw} for the sub-samples {\Be}, {\Bmu}, {\Bzeroe}, and {\Bzeromu} are shown in Table~\ref{table:ResultsdGdw_subsamples} and show 
very good consistency. The full correlation matrix of the systematic errors in different $w$-bins in the sub-sample results are determined with the 
approach described in Sect.~\ref{subsec:Systematics} and can be found in Ref.~\cite{epaps}. 
The weighted average of the differential rates is calculated by taking into account the full experimental correlations of all four individual 
measurements. The resulting central values, uncertainties and correlations are summarized in Table~\ref{table:ResultsdGdw}. 
Similarly, we calculate the branching fractions of the decays {\Be}, {\Bmu}, {\Bzeroe}, and {\Bzeromu} from the measured differential widths using 
the expression
\begin{equation}
 \mathcal{B} = \tau_B \sum\limits_i \Delta \Gamma_i ~.
\end{equation}
Here, $\tau_B$ is the corresponding $B$ meson lifetime and $\Delta \Gamma_i$ are the measured values of {\dGidw} times the $\Delta w$ used in the 
$i^\mathrm{th}$~bin.
The results are quoted in Table~\ref{table:BRResults}. We also quote combined results for charged and neutral $B$~meson decays and for all four 
sub-samples combined.
The ratio $R_D^{\mu/e} = \mathcal{B}(B \to D \mu \nu) / \mathcal{B}(B \to D e \nu)$ is found to be 
$0.995 \pm 0.022 (\mathrm{stat}) \pm 0.039 (\mathrm{syst})$.

\begin{table*}[htb]
\caption{The values of $\Delta\Gamma_i/\Delta w$ with the statistical and systematic uncertainties in the {\Be}, {\Bmu}, {\Bzeroe}, and {\Bzeromu} 
sub-samples. $i$, $w_{i,\mathrm{min}}$ and $w_{i,\mathrm{max}}$ are the $w$-bin number, lower and upper edge of the bin respectively. The value of 
$w_\mathrm{max}$ is 1.59209 for the sub-samples with a charged $B$ meson and 1.58901 for the sub-samples with a neutral $B$ meson. The 
$\Delta\Gamma_i/\Delta w$~results are statistically uncorrelated amongst bins and samples. The systematic correlations between bins and samples are 
given in Ref.~\cite{epaps}.}
\begin{tabular}
{@{\hspace{0.15cm}}l@{\hspace{0.15cm}} @{\hspace{0.15cm}}l@{\hspace{0.15cm}} @{\hspace{0.15cm}}l@{\hspace{0.15cm}} 
@{\hspace{0.15cm}}l@{\hspace{0.15cm}} @{\hspace{0.15cm}}l@{\hspace{0.15cm}} @{\hspace{0.15cm}}l@{\hspace{0.15cm}} 
@{\hspace{0.15cm}}l@{\hspace{0.15cm}} }
\hline \hline
 & & & \multicolumn{4}{c}{ \dGidw[$10^{-15}$GeV]  } \\ 
 $i$  & $w_{i,\mathrm{min}}$ & $w_{i,\mathrm{max}}$  & $B^{0} \to D^- e^+ \nu_e$ & $B^{0} \to D^- \mu^+ \nu_\mu$ & $B^+ \to \bar{D}^{0}e^+\nu_e$ & 
$B^+ \to \bar{D}^{0}\mu^+\nu_\mu$\\ \hline
0 & 1.00 & 1.06             &   $0.30 \pm 0.31 \pm 0.06$ &   $0.81 \pm 0.47 \pm 0.07$ &   $0.72 \pm 0.67 \pm 0.12$ &   $1.33 \pm 0.42 \pm 0.09$\\ 
1 & 1.06 & 1.12             &   $4.41 \pm 0.85 \pm 0.22$ &   $3.63 \pm 0.72 \pm 0.17$ &   $3.84 \pm 0.81 \pm 0.24$ &   $4.28 \pm 0.70 \pm 0.24$\\ 
2 & 1.12 & 1.18             &   $9.06 \pm 1.14 \pm 0.44$ &   $7.73 \pm 1.04 \pm 0.37$ &   $7.64 \pm 0.90 \pm 0.41$ &   $7.52 \pm 0.92 \pm 0.41$\\ 
3 & 1.18 & 1.24             &  $11.81 \pm 1.28 \pm 0.58$ &  $13.47 \pm 1.42 \pm 0.67$ &  $11.20 \pm 1.01 \pm 0.61$ &  $11.76 \pm 0.97 \pm 0.62$\\ 
4 & 1.24 & 1.30             &  $13.73 \pm 1.35 \pm 0.67$ &  $14.11 \pm 1.42 \pm 0.70$ &  $14.68 \pm 1.11 \pm 0.80$ &  $17.54 \pm 1.18 \pm 0.93$\\ 
5 & 1.30 & 1.36             &  $19.92 \pm 1.51 \pm 0.97$ &  $20.09 \pm 1.59 \pm 0.98$ &  $20.15 \pm 1.15 \pm 1.06$ &  $20.67 \pm 1.20 \pm 1.08$\\ 
6 & 1.36 & 1.42             &  $25.45 \pm 1.70 \pm 1.26$ &  $24.63 \pm 1.73 \pm 1.21$ &  $24.20 \pm 1.22 \pm 1.25$ &  $24.45 \pm 1.28 \pm 1.27$\\ 
7 & 1.42 & 1.48             &  $30.45 \pm 1.78 \pm 1.47$ &  $29.48 \pm 1.85 \pm 1.42$ &  $28.92 \pm 1.25 \pm 1.50$ &  $26.93 \pm 1.28 \pm 1.39$\\ 
8 & 1.48 & 1.54             &  $31.57 \pm 1.73 \pm 1.50$ &  $30.31 \pm 1.93 \pm 1.46$ &  $30.90 \pm 1.22 \pm 1.57$ &  $29.85 \pm 1.36 \pm 1.50$\\ 
9 & 1.54 & $w_\mathrm{max}$ &  $35.81 \pm 1.88 \pm 1.68$ &  $34.62 \pm 2.19 \pm 1.63$ &  $34.42 \pm 1.24 \pm 1.73$ &  $32.83 \pm 1.44 \pm 1.63$\\ 
\hline \hline
\end{tabular}
\label{table:ResultsdGdw_subsamples}
\end{table*}
\begin{table*}[htb]
\caption{The values of $\Delta\Gamma_i/\Delta w$ obtained in different bins of $w$ after combination of the {\Be}, {\Bmu}, {\Bzeroe}, and {\Bzeromu} 
sub-samples. The columns are (from left to right) the bin number, the lower and the upper edge of the $i^\mathrm{th}$~bin, the value of 
$\Delta\Gamma_i/\Delta w$ in this bin with the statistical and systematic uncertainties, and the correlation matrix of the systematic error. 
The value of $w_\mathrm{max} = 1.59055$ is the average of the values for charged and neutral $B$ mesons.}
\begin{tabular}
{@{\hspace{0.15cm}}l@{\hspace{0.15cm}} @{\hspace{0.15cm}}l@{\hspace{0.15cm}} @{\hspace{0.15cm}}l@{\hspace{0.15cm}} 
@{\hspace{0.15cm}}l@{\hspace{0.15cm}} @{\hspace{0.15cm}}l@{\hspace{0.15cm}} @{\hspace{0.15cm}}l@{\hspace{0.15cm}} 
@{\hspace{0.15cm}}l@{\hspace{0.15cm}} @{\hspace{0.15cm}}l@{\hspace{0.15cm}} @{\hspace{0.15cm}}l@{\hspace{0.15cm}} 
@{\hspace{0.15cm}}l@{\hspace{0.15cm}} @{\hspace{0.15cm}}l@{\hspace{0.15cm}} @{\hspace{0.15cm}}l@{\hspace{0.15cm}} 
@{\hspace{0.15cm}}l@{\hspace{0.15cm}} @{\hspace{0.15cm}}l@{\hspace{0.15cm}} }
\hline \hline
 & & & & \multicolumn{10}{c}{ $\rho_{ij,\mathrm{syst}}$} \\ 
$i$  & $w_{i,\mathrm{min}}$ & $w_{i,\mathrm{max}}$  & \dGidw [$10^{-15}$GeV]    & 0 & 1 & 2 & 3 & 4 & 5 & 6 & 7 & 8 & 9\\ \hline
0 & 1.00 & 1.06             &  $ 0.68 \pm 0.21 \pm 0.05 $  & 1.000 & 0.682 & 0.677 & 0.663 & 0.654 & 0.656 & 0.664 & 0.648 & 0.608 & 0.560\\
1 & 1.06 & 1.12             &  $ 3.88 \pm 0.38 \pm 0.18 $  &       & 1.000 & 0.976 & 0.974 & 0.969 & 0.972 & 0.972 & 0.961 & 0.933 & 0.900\\
2 & 1.12 & 1.18             &  $ 7.59 \pm 0.50 \pm 0.35 $  &       &       & 1.000 & 0.991 & 0.987 & 0.990 & 0.989 & 0.980 & 0.959 & 0.929\\
3 & 1.18 & 1.24             & $ 11.42 \pm 0.58 \pm 0.54 $  &       &       &       & 1.000 & 0.993 & 0.993 & 0.990 & 0.980 & 0.961 & 0.934\\
4 & 1.24 & 1.30             & $ 14.59 \pm 0.64 \pm 0.69 $  &       &       &       &       & 1.000 & 0.996 & 0.992 & 0.985 & 0.972 & 0.952\\
5 & 1.30 & 1.36             & $ 19.49 \pm 0.69 \pm 0.91 $  &       &       &       &       &       & 1.000 & 0.996 & 0.991 & 0.979 & 0.956\\
6 & 1.36 & 1.42             & $ 23.66 \pm 0.76 \pm 1.10 $  &       &       &       &       &       &       & 1.000 & 0.995 & 0.981 & 0.952\\
7 & 1.42 & 1.48             & $ 27.56 \pm 0.79 \pm 1.27 $  &       &       &       &       &       &       &       & 1.000 & 0.992 & 0.968\\
8 & 1.48 & 1.54             & $ 29.52 \pm 0.80 \pm 1.34 $  &       &       &       &       &       &       &       &       & 1.000 & 0.985\\
9 & 1.54 & $w_\mathrm{max}$ & $ 33.37 \pm 0.86 \pm 1.50 $  &       &       &       &       &       &       &       &       &       & 1.000\\
\hline \hline
\end{tabular}
\label{table:ResultsdGdw}
\end{table*}
\begin{table}[htb]
\caption{Branching fractions of the decays {\Be}, {\Bmu}, {\Bzeroe}, and {\Bzeromu}. The branching fractions of {\Bl} ({\Bzerol}) are the weighted 
averages of the {\Be} and {\Bmu} ({\Bzeroe} and {\Bzeromu}) branching fraction results. The last row of the table corresponds to the branching 
fraction of all four sub-samples combined, expressed in terms of the neutral mode {\Bzerol} assuming the lifetime $\tau_{B^0} 
= 1.519$~\cite{Agashe:2014kda}. The first error on the yields and on the branching fractions is statistical. The second uncertainty is systematic.}
\begin{tabular}
 {@{\hspace{0.2cm}}l@{\hspace{0.2cm}} @{\hspace{0.2cm}}l@{\hspace{0.2cm}} @{\hspace{0.2cm}}l@{\hspace{0.2cm}}}
\hline \hline
Sample & Signal yield & $\mathcal{B}$ [\%] \\ 
\hline
$B^{0} \to D^- e^+ \nu_e$           & $2848 \pm 72 \pm 17$    & $2.44 \pm 0.06 \pm 0.12$\\ 
$B^{0} \to D^- \mu^+ \nu_\mu$       & $2302 \pm 63 \pm 13$    & $2.39 \pm 0.06 \pm 0.11$\\ 
$B^+ \to \bar{D}^{0}e^+\nu_e$       & $6456 \pm 126 \pm 66$   & $2.57 \pm 0.05 \pm 0.13$\\ 
$B^+ \to \bar{D}^{0}\mu^+\nu_\mu$   & $5386 \pm 110 \pm 51$   & $2.58 \pm 0.05 \pm 0.13$\\ 
\hline
$B^{0} \to D^- \ell^+ \nu_\ell$     & $5150 \pm 95 \pm 29$    & $2.39 \pm 0.04 \pm 0.11$\\ 
$B^+ \to \bar{D}^{0}\ell^+\nu_\ell$ & $11843 \pm 167 \pm 120$ & $2.54 \pm 0.04 \pm 0.13$\\ 
\hline
$B \to D\ell\nu_\ell$               & $16992 \pm 192 \pm 142$ & $2.31 \pm 0.03 \pm 0.11$\\ 
\hline \hline
\end{tabular}
\label{table:BRResults}
\end{table}

\subsection{Systematic uncertainties}
\label{subsec:Systematics}

We use a toy MC approach to estimate systematic uncertainties of the values of $\Delta\Gamma_i/\Delta w$ and their correlations. For a given 
systematic error component, we vary one or several parameters in the MC simulation according to a Gaussian distribution with a width corresponding to 
the systematic uncertainty under study. This altered MC sample is then used to repeat the entire analysis procedure, resulting in an updated value of 
$\Delta\Gamma_i/\Delta w$. Repeating this procedure 1000 times, we obtain a distribution of $\Delta\Gamma_i/\Delta w$ values corresponding to this 
specific systematic error component. The distribution is fitted with a Gaussian function and the width $\sigma_i$ of the Gaussian function is taken as 
the estimate of the contribution of this error component to the total systematic uncertainty. The corresponding correlation $\rho_{i,j}$ between 
{\dGidw} and {\dGjdw} is calculated as
\begin{equation}
\rho_{i,j}=\frac
{ \langle (\dGidwf - \langle \dGidwf \rangle) (\dGjdwf - \langle \dGjdwf \rangle) \rangle } 
{ \sqrt{\langle(\dGidwf - \langle \dGidwf \rangle)^2\rangle} \sqrt{\langle(\dGjdwf - \langle \dGjdwf \rangle)^2\rangle} }~,
\end{equation}
where the average indicated by the brackets is taken over the toy MC sample. To reduce the effect of outliers, toy MC events where one value of 
{\dGidw} lies outside of the interval $\pm 3\sigma_i$ are removed. The elements of the covariance matrix are then calculated as 
$\rho_{i,j}\sigma_i\sigma_j$. The full systematic error matrix is obtained by adding the covariance matrices corresponding to the individual error 
components linearly. This is equivalent to the quadratic addition of the systematic error components of {\dGidw}. The individual systematic error 
components are described in the following.

{\it Tag correction}: This error component is estimated in two steps: we apply all the corrections to the MC mentioned in Sect.~\ref{subsec:Data 
Sample and Event Selection} and vary these within their respective uncertainties. This results in systematic uncertainties in the 480 tag correction 
coefficients introduced in Sect.~\ref{subsec:Hadronic Tagging}. Finally, we propagate the uncertainties in the tag correction coefficients to the 
values of {\dGidw}. The statistical uncertainties in the tag corrections are varied independently while the systematic errors on the coefficients are 
conservatively assumed to be 100\% correlated.

{\it Charged track reconstruction}: We assume a 0.35\% reconstruction uncertainty for each charged particle in the final state. This uncertainty is 
added linearly for each charged particle on the \emph{signal side}, as the charged particle reconstruction on the tag side is already corrected by our 
tag calibration. This uncertainty is propagated to {\dGidw} using the toy MC approach.

{\it Branching fractions and form factors (FF)}: We adjust the branching fraction and the CLN form factor of the decay {\BDstarlnu} -- the main 
cross-feed background -- in the MC~\cite{Agashe:2014kda,HFAG_averages}. Also, for semileptonic decays to orbitally excited $D$~meson states 
{\BDstarstarlnu}, we correct both the rate and the form factor~\cite{Agashe:2014kda,leibovich1998semileptonic}. For the $D$~meson decays, only the 
branching fractions are adjusted~\cite{Agashe:2014kda}. The error component corresponding to charmless semileptonic decays {\BXulnu} 
contains both the uncertainty in the inclusive $b\to u \ell \nu$ rate~\cite{HFAG_averages} and in the known exclusive 
decays ($B\to\pi\ell\nu,\rho\ell\nu,\omega\ell\nu,\eta\ell\nu,\eta'\ell\nu$)~\cite{Agashe:2014kda}.

{\it Signal shape}: This error component corresponds to the uncertainty in the smearing parameter of the signal shape correction described in Sect.~\ref{subsec:Yield extraction}.

{\it $B$~lifetime}: The lifetimes of $B^0$ and $B^+$ are needed in Eq.~(\ref{eq:dGdw}) to determine {\dGidw}. We use the following 
central values and uncertainties: $\tau(B^0)=1.519\pm 0.005$~ps and $\tau(B^+)=1.638\pm 0.004$~ps~\cite{Agashe:2014kda}.

{\it Particle identification}: Due to the use of the tag calibration sample, the uncertainty in the charged lepton identification cancels. A 
remaining particle-identification uncertainty arises from kaon and pion identification, which is estimated using a data sample of $D^{*+}\to D^0\pi^+$ 
decays. The misidentification probability of pions as electrons or as muons is also adjusted in MC simulation by using real $D^{*+}\to D^0\pi^+$ 
events.

{\it Luminosity}: This component includes the uncertainty in the measurement of the Belle data luminosity (1.4\%) and the uncertainty in the 
branching fraction of $\Upsilon(4S)\to B\bar B$~\cite{Agashe:2014kda}. The luminosity measurement uses Bhabha events and its uncertainty is 
dominated by the accuracy of the event generator used.

The systematic uncertainties in {\dGidw} are itemized in Table~\ref{table:sysErrs}. Since signal is suppressed at zero recoil, the zeroth bin has the 
largest relative uncertainty. The systematic uncertainties of the branching fractions in Table~\ref{table:BRResults} are estimated by using the same 
toy MC approach and the same error components.
\begin{table}[h!tb]
\caption{Itemization of the systematic uncertainty in {\dGidw} in each $w$ bin. Refer to the main text for more details on the systematic error 
components.}
\begin{tabular} 
{@{\hspace{0.1cm}}l@{\hspace{0.1cm}}  @{\hspace{0.1cm}}l@{\hspace{0.1cm}} @{\hspace{0.1cm}}l@{\hspace{0.1cm}} @{\hspace{0.1cm}}l@{\hspace{0.1cm}} 
@{\hspace{0.1cm}}l@{\hspace{0.1cm}} @{\hspace{0.1cm}}l@{\hspace{0.1cm}} @{\hspace{0.1cm}}l@{\hspace{0.1cm}} @{\hspace{0.1cm}}l@{\hspace{0.1cm}} 
@{\hspace{0.1cm}}l@{\hspace{0.1cm}} @{\hspace{0.1cm}}l@{\hspace{0.1cm}} @{\hspace{0.1cm}}l@{\hspace{0.1cm}}}
\hline \hline 
 & \multicolumn{10}{c}{ $\sigma \left( \dGidw \right) $[\%]  } \\ 
 & 0 & 1 & 2 & 3 & 4 & 5 & 6 & 7 & 8 & 9\\ \hline 
Tag correction & 3.0 & 3.2 & 3.3 & 3.4 & 3.4 & 3.4 & 3.4 & 3.3 & 3.3 & 3.2\\
Charged tracks & 1.7 & 1.6 & 1.6 & 1.6 & 1.6 & 1.6 & 1.6 & 1.6 & 1.6 & 1.6\\
$\mathcal{B}( D \to $ hadronic $)$ & 2.0 & 1.8 & 1.8 & 1.8 & 1.8 & 1.8 & 1.8 & 1.9 & 1.9 & 1.9\\
$\mathcal{B}( B \to D^{*(*)} \ell \nu )$ & 1.3 & 0.8 & 0.8 & 0.9 & 0.8 & 0.7 & 0.5 & 0.2 & 0.2 & 0.4\\
$\mathcal{B}( B \to X_u \ell \nu )$ & 0.4 & 0.1 & 0.0 & 0.1 & 0.0 & 0.0 & 0.0 & 0.0 & 0.0 & 0.0\\
FF($B \to D^{*} \ell \nu$) & 0.4 & 0.2 & 0.2 & 0.2 & 0.2 & 0.1 & 0.1 & 0.1 & 0.1 & 0.2\\
FF($B \to D^{**} \ell \nu$) & 2.5 & 1.2 & 0.9 & 0.7 & 0.5 & 0.5 & 0.7 & 0.5 & 0.1 & 0.4\\
Signal shape & 5.0 & 0.8 & 0.6 & 0.5 & 0.5 & 0.4 & 0.3 & 0.3 & 0.2 & 0.1\\
Lifetimes & 0.2 & 0.2 & 0.2 & 0.2 & 0.2 & 0.2 & 0.2 & 0.2 & 0.2 & 0.2\\
$\pi^0$ efficiency & 0.9 & 0.6 & 0.6 & 0.6 & 0.6 & 0.6 & 0.6 & 0.6 & 0.6 & 0.7\\
$K/\pi$ efficiency & 1.1 & 0.9 & 0.9 & 0.9 & 0.9 & 0.9 & 0.9 & 1.0 & 1.0 & 1.0\\
$K_S$ efficiency & 0.4 & 0.2 & 0.2 & 0.2 & 0.2 & 0.2 & 0.2 & 0.2 & 0.2 & 0.2\\
Luminosity & 1.4 & 1.4 & 1.5 & 1.4 & 1.4 & 1.4 & 1.4 & 1.4 & 1.4 & 1.4\\\hline
Total & 7.3 & 4.7 & 4.7 & 4.7 & 4.7 & 4.6 & 4.7 & 4.6 & 4.5 & 4.5\\
\hline \hline 
\end{tabular} 
\label{table:sysErrs}
\end{table}

\section{Discussion}
\label{sec:Discussion}

\subsection{CLN parameterization interpretation}
\label{subsec:CLN}

The usual approach used in the literature~\cite{HFAG_averages} to interpret the $\Delta\Gamma/\Delta w$ distribution is to perform a fit to the CLN 
form-factor parameterization (Eq.~\ref{eq:CLN}), determine {\VGE} and obtain {\VE} by dividing by {\Gone}. We do so here and determine the overall 
normalization {\VGE} and the parameter $\rho^2$ of the CLN form-factor parameterization by minimizing the $\chi^2$~function
\begin{equation}
  \chi^2 = \sum\limits_{i,j}(\dGidwf-\dGiCLNdwf)\mathbf{C}^{-1}_{ij}(\dGjdwf-\dGjCLNdwf)~,
\end{equation}
where {\dGidw} is the measured value from Table~\ref{table:ResultsdGdw_subsamples}~or~\ref{table:ResultsdGdw} and {\dGiCLNdw} 
is the partial width calculated using Eqs.~\ref{eq:dGammadw} and \ref{eq:CLN}:
\begin{equation}
  \dGiCLNdwf(\VGE,\rho^2)=\frac{1}{\Delta w}\int_{w_{i,\mathrm{min}}}^{w_{i,\mathrm{max}}} \frac{d \Gamma_{\mathrm{CLN}}}{dw} dw~.
\end{equation}
The total covariance matrix $\mathbf{C}$ is the sum of the diagonal statistical error matrix $\mathbf{C}_\mathrm{stat}$ and the systematic covariance 
matrix $\mathbf{C}_\mathrm{syst}$, calculated from the systematic errors and correlations presented in Sect.~\ref{sec:results}.
For the fit on the combined sample we use the averaged nominal masses of charged and neutral mesons ($m_B=5.27942$~GeV and $m_D=1.86723$~GeV).

\begin{figure}
\includegraphics[width=\columnwidth]{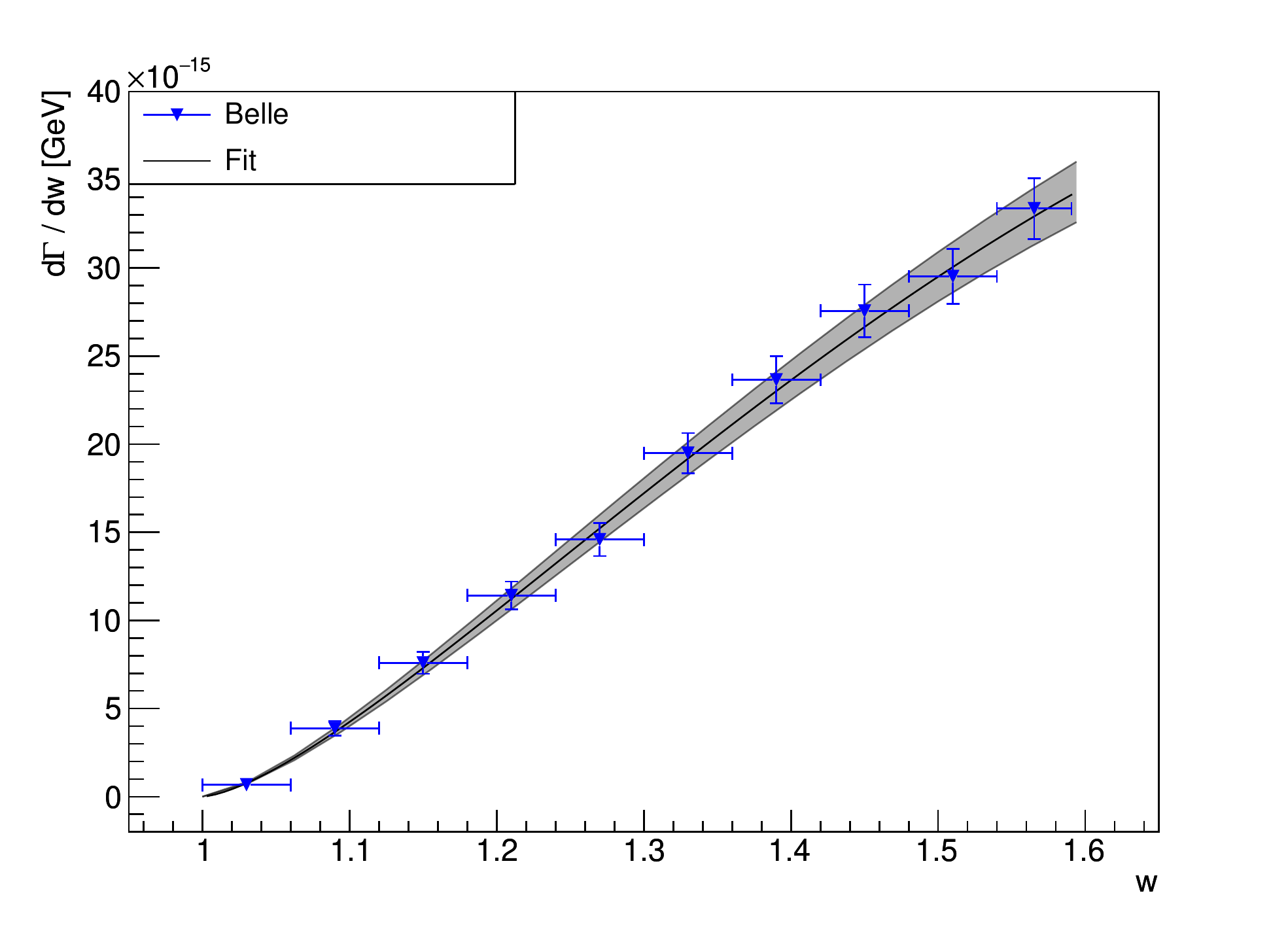}
\caption{Fit to the measured $\Delta\Gamma/\Delta w$ spectrum of the decay {\BDlnu}, assuming the CLN form-factor parameterization 
(Eq.~(\ref{eq:CLN})). The points with error bars are the data. Their respective uncertainties are shown by the vertical error bars; the bin widths 
are shown by the horizontal bars. The solid curve corresponds to the result of the fit. The shaded area around this curve indicates the uncertainty in 
the coefficients of the CLN parameters.} 
\label{fig:w_fit}
\end{figure}

\begin{figure}[h!t]
\centering
\includegraphics[width=\columnwidth]{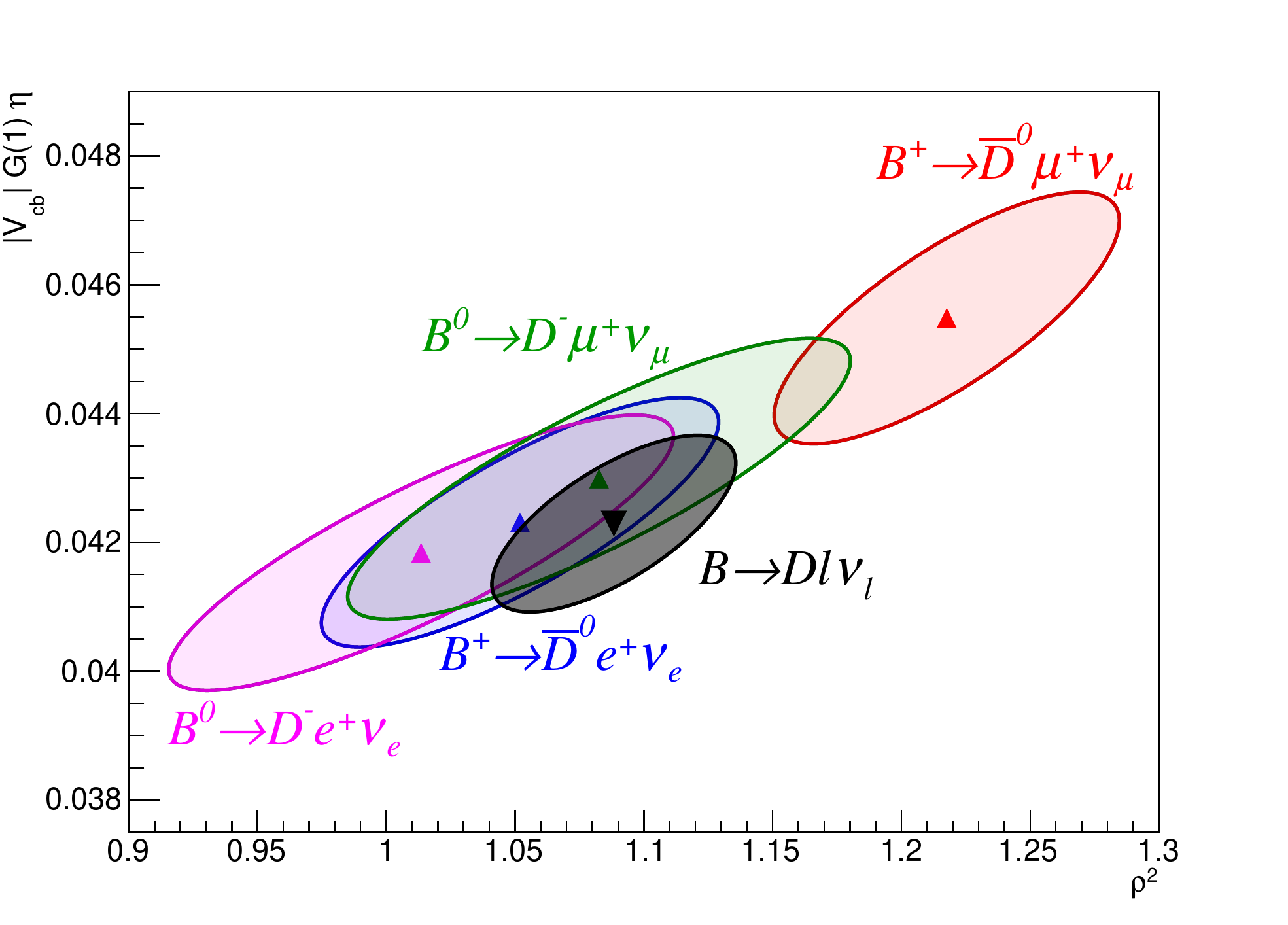}
\caption{Result of the fit assuming the CLN form-factor parameterization (Eq.~(\ref{eq:CLN})). The error ellipses ($\Delta\chi^2=1$) of {\VGE} and 
$\rho^2$ are shown for the fit to the {\Be}, {\Bmu}, {\Bzeroe}, and {\Bzeromu} sub-samples, and to the combined sample.} 
\label{fig:ResultEllipses}
\end{figure}

The result of the fit is shown in Fig.~\ref{fig:w_fit}. The results in terms of {\VGE} and $\rho^2$ are given in Table~\ref{table:Results} and 
Fig.~\ref{fig:ResultEllipses}, separately for the {\Be}, {\Bmu}, {\Bzeroe}, and {\Bzeromu} sub-samples and for the combined spectrum. Assuming the 
form-factor normalization {\Gone} derived in Ref.~\cite{lattice2015b}
\begin{equation}
 \Gone=1.0541\pm 0.0083~, \label{eq:gone_detar}
\end{equation}
we obtain $\VE=(40.12\pm 1.34)\times 10^{-3}$.
\begin{table*}[h!tb]
\caption{Result of the fit to the measured $\Delta\Gamma/\Delta w$ spectrum of the decay {\BDlnu} using the CLN form-factor parameterization 
(Eq.~(\ref{eq:CLN})). The CLN parameters {\VGE} and $\rho^2$ are given for the {\Be}, {\Bmu}, {\Bzeroe}, and {\Bzeromu} sub-samples and for all four 
sub-samples combined (based on the combined sample shown in Table~\ref{table:ResultsdGdw}). The value of {\VE} is obtained assuming the form-factor 
normalization in Eq.~(\ref{eq:gone_detar}). ``Correlation'' denotes the 
measured correlation between the overall uncertainties of {\VGE} and $\rho^2$. }
\begin{tabular}
 {@{\hspace{0.15cm}}l@{\hspace{0.15cm}} @{\hspace{0.15cm}}l@{\hspace{0.15cm}} @{\hspace{0.15cm}}l@{\hspace{0.15cm}} 
  @{\hspace{0.15cm}}l@{\hspace{0.15cm}} @{\hspace{0.15cm}}l@{\hspace{0.15cm}} | @{\hspace{0.15cm}}l@{\hspace{0.15cm}} }
\hline \hline
  & $B^+ \to \bar{D}^{0}e^+\nu_e$ & $B^+ \to \bar{D}^{0}\mu^+\nu_\mu$ & $B^{0} \to D^- e^+ \nu_e$ & $B^{0} \to D^- \mu^+ \nu_\mu$ & $B \to 
D\ell\nu_\ell$\\ \hline
 \VGE [$10^{-3}$]  & $ 42.31 \pm 1.94$  & $ 45.48 \pm 1.96$  & $ 41.84 \pm 2.14$  & $ 42.99 \pm 2.18$  & $ 42.29 \pm 1.37$ \\ 
 $\rho^2$  & $ 1.05 \pm 0.08$  & $ 1.22 \pm 0.07$  & $ 1.01 \pm 0.10$  & $ 1.08 \pm 0.10$  & $ 1.09 \pm 0.05$ \\ 
 Correlation  & $ 0.81$  & $ 0.77$  & $ 0.85$  & $ 0.84$  & $ 0.69$ \\ 
\hline 
 \VE [$10^{-3}$]  & $ 40.14 \pm 1.86$  & $ 43.15 \pm 1.89$  & $ 39.69 \pm 2.05$  & $ 40.78 \pm 2.09$  & $ 40.12 \pm 1.34$ \\ 
\hline 
$\chi^2 / n_\mathrm{df}$ & $ 2.19/8$  & $ 2.71/8$  & $ 9.65/8$  & $ 4.36/8$  & $ 4.57/8$ \\ 
Prob. & $ 0.97$  & $ 0.95$  & $ 0.29$  & $ 0.82$  & $ 0.80$ \\ 
\hline \hline
\end{tabular}
\label{table:Results}
\end{table*}

\subsection{Model-independent BGL fit}

Recent lattice data at non-zero recoil~\cite{lattice2015b,na2015b} allows us to perform a combined fit to the BGL form factor. We proceed as in the 
previous section and minimize the $\chi^2$ function
\begin{equation}
  \chi^2 = \sum\limits_{i,j}(\dGidwf-\dGiBGLdwf)\mathbf{C}^{-1}_{ij}(\dGjdwf-\dGjBGLdwf) 
  + \sum\limits_{k,l}\left(f^\mathrm{LQCD}_{+,0}(w_k)-f^\mathrm{BGL}_{+,0}(w_k)\right)\mathbf{D}^{-1}_{kl}
  \left((f^\mathrm{LQCD}_{+,0} (w_l)-f^\mathrm {BGL}_{+,0}(w_l)\right)~.
\end{equation}
Again, {\dGidw} is taken from Table~\ref{table:ResultsdGdw} and {\dGiBGLdw} is the partial width calculated using Eqs.~\ref{eq:dGammadw}, 
\ref{eq:fplusAndG}, and \ref{eq:BGL}
\begin{equation}
  \dGiBGLdwf(\VE,a_{+,n})=\frac{1}{\Delta w}\int_{w_{i,\mathrm{min}}}^{w_{i,\mathrm{max}}} \frac{d \Gamma_{\mathrm{BGL}}}{dw} 
dw~.
\end{equation}
The error matrix $\mathbf{C}$ includes the statistical and systematic uncertainties in the measurements of $\Delta\Gamma_i/\Delta w$. The data is fit 
together with predictions of lattice QCD (LQCD), which are available for the form factors $f_+(w)$ and $f_0(w)$ at selected points in $w$. The second 
sum runs over all LQCD predictions included in the fit and the corresponding error matrix $\mathbf{D}$ contains the LQCD uncertainty in these 
predictions. 
We use lattice data obtained by the FNAL/MILC and HPQCD collaborations~\cite{lattice2015b,na2015b}. 
Both LQCD calculations are dominated by their systematic errors. The correlation between them is expected to be small since the collaborations use different heavy-quark methods, 
lattice NRQCD \cite{Lepage:1992tx} for HPQCD and the Fermilab method \cite{ElKhadra:1996mp} for FNAL/MILC. 
We therefore assume the two LQCD results to be uncorrelated in our fits.

Note that LQCD yields results for both the $f_+$ and $f_0$ form factors while the experimental distribution $\Delta\Gamma_i/\Delta w$ 
depends on $f_+$ only. Using the kinematic constraint from Eq.~\ref{eq:kinematic}, we can include the LQCD results for $f_0$ into the fit, allowing us 
to better constrain $f_+$. Following Ref.~\cite{lattice2015b}, we implement this constraint by expressing $a_{0,0}$ in terms of the other $a_{+,n}$ 
and $a_{0,n}$ coefficients.
FNAL/MILC obtains values for both the $f_+$ and the $f_0$ form factors at $w$ values of 1, 1.08, and 1.16. The full covariance matrix for these six measurements is available in 
Table~VII of Ref.~\cite{lattice2015b}.

The form factors determined by HPQCD are based on a different form factor parameterization by Bourrely, Caprini and Lellouch (BCL), see Ref.~\cite{BCL}.
BCL uses an expansion in a conformal mapping variable to offer perturbative QCD scaling also at higher $q^2$ values. 
The formulae and pole choices used by HPQCD can be seen in Eqs.~A1 to A6 of Ref.~\cite{na2015b}. 
As a result of their fit they provide the coefficients $a^{(0)}_0$, $a^{(0)}_1$, $a^{(0)}_2$, $a^{(+)}_0$, $a^{(+)}_1$, and $a^{(+)}_2$, together with their $6\times 6$ covariance 
matrix (Table VII of Ref.~\cite{na2015b}). To be able to include these results in the same fit as the FNAL/MILC points, we transform the coefficients into the form-factor 
values of $f_+$ and $f_0$ at $w=1, 1.08$ and $1.16$:
\begin{equation}
 \left(\begin{array}{c}
        f_0(1)\\
        f_0(1.08)\\
        f_0(1.16)\\
        f_+(1)\\
        f_+(1.08)\\
        f_+(1.16)
       \end{array}\right)=
       \mathbf{M}
       \left(\begin{array}{c}
        a^{(0)}_0 \\
        a^{(0)}_1 \\
        a^{(0)}_2 \\
        a^{(+)}_0 \\
        a^{(+)}_1 \\
        a^{(+)}_2
       \end{array}\right)~,
\end{equation}
where $\mathbf{M}$ is a block-diagonal $6\times 6$ matrix. Denoting the covariance matrix of the HPQCD $a$-parameters by $\mathbf{Cov}(a)$, the error 
matrix of the form-factor values becomes $\mathbf{M}~\mathbf{Cov}(a)~\mathbf{M}^{T}$. The HPQCD results in terms of the $f_+$ and $f_0$ form factors 
at $w=1, 1.08$ and $1.16$, together with their correlation coefficients, are given in Table~\ref{table:HPQCD_f}.

\begin{table*}[htb]
\caption{Lattice QCD results obtained by the HPQCD collaboration~\cite{na2015b}, expressed in terms of $f_+$ and $f_0$~form-factor values at $w=1, 
1.08$ and $1.16$.} \label{table:HPQCD_f}
\begin{tabular} 
{ @{\hspace{0.2cm}}l@{\hspace{0.2cm}} | @{\hspace{0.2cm}}c@{\hspace{0.2cm}} | @{\hspace{0.2cm}}c@{\hspace{0.2cm}} @{\hspace{0.2cm}}c@{\hspace{0.2cm}} 
  @{\hspace{0.2cm}}c@{\hspace{0.2cm}} @{\hspace{0.2cm}}c@{\hspace{0.2cm}} @{\hspace{0.2cm}}c@{\hspace{0.2cm}} @{\hspace{0.2cm}}c@{\hspace{0.2cm}}}
\hline \hline 
 & & \multicolumn{6}{c}{Correlation coefficients}\\
		& Central value 	& $f_+(1)$ 	& $f_+(1.08)$ 	& $f_+(1.16)$ 	& $f_0(1)$ 	& $f_0(1.08)$ 	& $f_0(1.16)$ 	\\ 
\hline
$f_+(1)$	&$1.178 \pm 0.046$	& 1.000		&	0.989	&	0.954	&	0.507	&	0.518	&	0.525 \\
$f_+(1.08)$	&$1.082 \pm 0.041$	&		&	1.000	&	0.988	&	0.582	&	0.600	&	0.615 \\
$f_+(1.16)$	&$0.996 \pm 0.037$	&		&		&	1.000	&	0.650	&	0.676	&	0.698 \\
$f_0(1)$	&$0.902 \pm 0.041$	&		&		&		&	1.000	&	0.995	&	0.980 \\
$f_0(1.08)$	&$0.860 \pm 0.038$	&		&		&		&		&	1.000	&	0.995 \\
$f_0(1.16)$	&$0.821 \pm 0.036$	&		&		&		&		&		&	1.000 \\
\hline \hline 
\end{tabular}
\end{table*}

\begin{figure}[ht]
\centering
\includegraphics[width=\columnwidth]{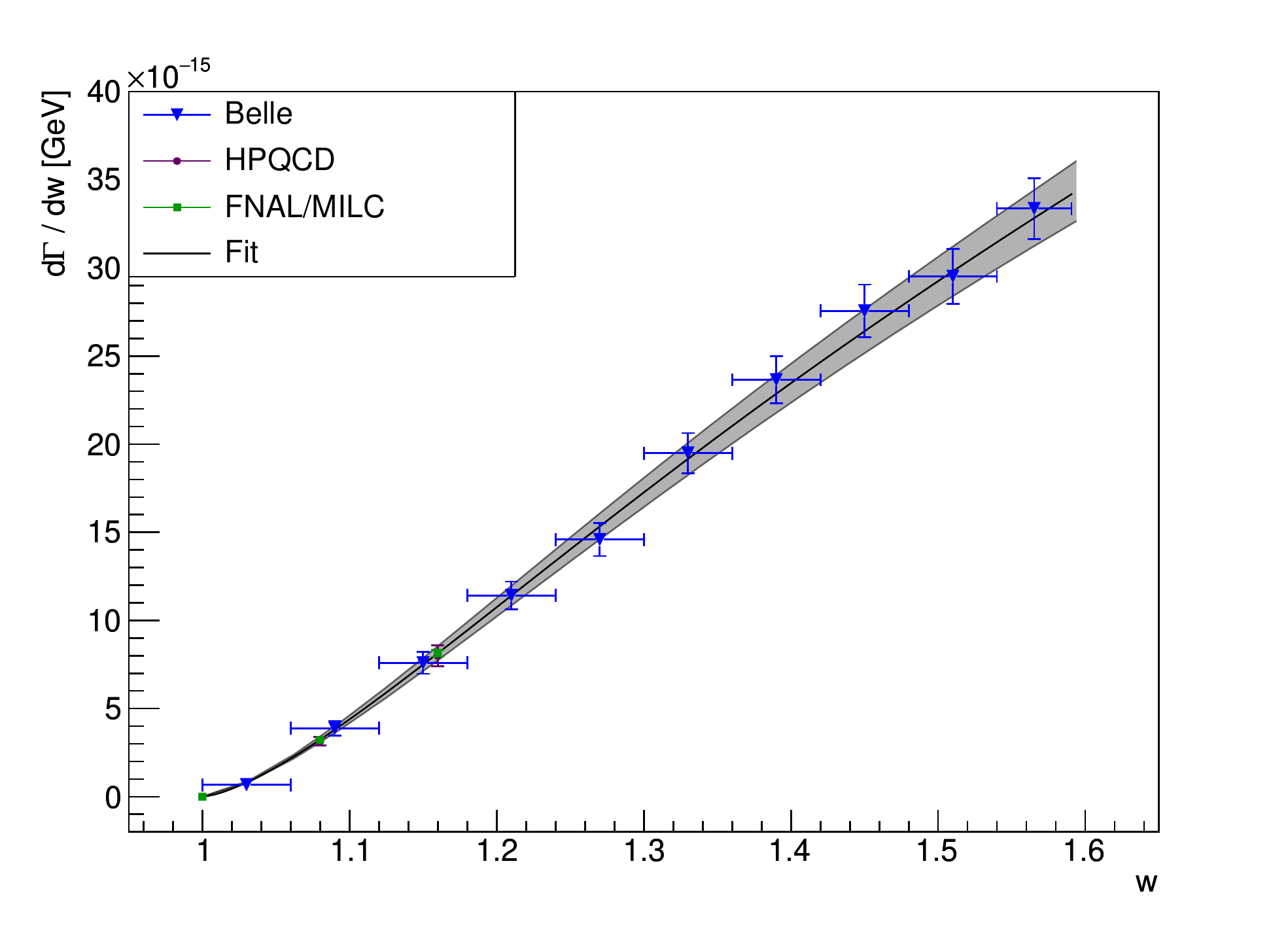}
\caption{Differential width of ${\BDlnu}$ and result of the combined fit to experimental and lattice QCD (FNAL/MILC and HPQCD) data. The BGL series 
(Eq.~(\ref{eq:BGL})) is truncated after the cubic term. The points with error bars are Belle and LQCD data (only results for $f_+$ are shown on this 
plot). For Belle data, the uncertainties are represented by the vertical error bars and the bin widths by the horizontal bars. The solid curve 
corresponds to the result of the fit. The shaded area around this curve indicates the uncertainty in the coefficients of the BGL series.}
\label{fig:MIFit_dGdw}
\end{figure}

\begin{figure}[ht]
\centering
\includegraphics[width=\columnwidth]{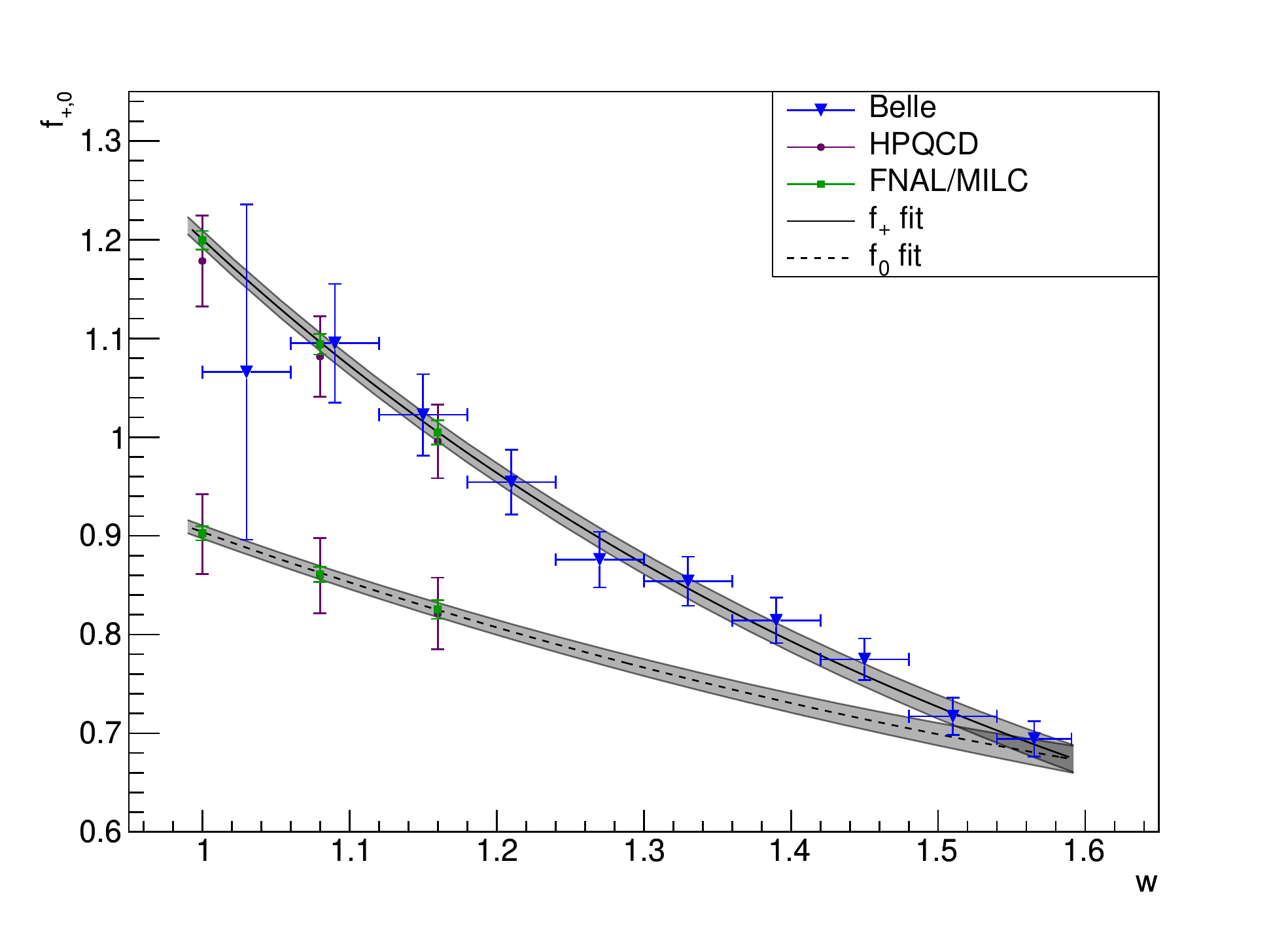}
\caption{Form factors of the decay {\BDlnu} and result of the combined fit to experimental and lattice QCD (FNAL/MILC and HPQCD) data. The BGL series 
(Eq.~(\ref{eq:BGL})) is truncated after the cubic term. The points with error bars are Belle and LQCD data. The solid curve is the $f_+$ form factor 
and the dashed curve represents $f_0$. The shaded areas around these curves indicate the uncertainty in the coefficients of the BGL expansion.}
\label{fig:MIFit_f}
\end{figure}

Table~\ref{table:MIFit_dGdw} shows the result of the BGL fit to experimental and LQCD data (FNAL/MILC and HPQCD) for different truncation orders of 
the series ($N=2,3,4$). To implement the unitarity bound (Eq.~(\ref{eq:unitarity})), we constrain the cubic and quartic coefficients in 
Eq.~(\ref{eq:BGL}) to $0\pm 1$ in the fits with $N=3$ and $N=4$ by adding measurement points of $a_{+,i\geq3}$ and $a_{0,i\geq3}$ to the $\chi^2$.
This follows the method in Ref.~\cite{lattice2015b} and results in a constant number of degrees of freedom. 
For $N\geq3$, the fit stabilizes and we get a reasonable goodness of fit. We thus 
choose this truncation order as our preferred fit. The fit result in terms of $\Delta\Gamma/\Delta w$ and $f_{+,0}$ is shown for $N=3$ in 
Figs.~\ref{fig:MIFit_dGdw} and \ref{fig:MIFit_f}, respectively. Our baseline result for {\VE} for the combined fit to experimental and lattice QCD 
data is thus $(41.10 \pm 1.14)\times 10^{-3}$. This is slightly more precise than the fit result using the CLN form-factor parameterization (2.8\% 
vs. 3.3\%) due to the additional input from LQCD. 
The additional lattice points are also the dominant cause of differences in the resulting values.
We have verified the stability of this {\VE} value by repeating the fit with different sets of 
lattice QCD data (Table~\ref{table:MIFit_comparison}) and the differences between the results are well below one standard deviation.
\begin{table}[htb]
\caption{Result of the combined fit to experimental and lattice QCD (FNAL/MILC and HPQCD) data for different truncation orders of the BGL series 
(Eq.~(\ref{eq:BGL})). Note that the value of $a_{0,0}$ is not determined from the fit but rather inferred using the kinematic constraint 
(Eq.~(\ref{eq:kinematic})).}
\begin{tabular} 
{ @{\hspace{0.1cm}}l@{\hspace{0.1cm}} @{\hspace{0.1cm}}l@{\hspace{0.1cm}} @{\hspace{0.1cm}}l@{\hspace{0.1cm}} @{\hspace{0.1cm}}l@{\hspace{0.1cm}} }
\hline \hline 
 & $N = 2$ & $N = 3$ & $N = 4$\\ \hline 
$a_{+,0}$ & 0.0127 $\pm$ 0.0001 & 0.0126 $\pm$ 0.0001 & 0.0126 $\pm$ 0.0001\\ 
$a_{+,1}$ & -0.091 $\pm$ 0.002 & -0.094 $\pm$ 0.003 & -0.094 $\pm$ 0.003\\ 
$a_{+,2}$ & 0.34 $\pm$ 0.03 & 0.34 $\pm$ 0.04 & 0.34 $\pm$ 0.04\\ 
$a_{+,3}$ &  --  & -0.1 $\pm$ 0.6 & -0.1 $\pm$ 0.6\\ 
$a_{+,4}$ &  --  &  --  & 0.0 $\pm$ 1.0\\ 
\hline 
$a_{0,0}$ & 0.0115 $\pm$ 0.0001 & 0.0115 $\pm$ 0.0001 & 0.0115 $\pm$ 0.0001\\ 
$a_{0,1}$ & -0.058 $\pm$ 0.002 & -0.057 $\pm$ 0.002 & -0.057 $\pm$ 0.002\\ 
$a_{0,2}$ & 0.22 $\pm$ 0.02 & 0.12 $\pm$ 0.04 & 0.12 $\pm$ 0.04\\ 
$a_{0,3}$ &  --  & 0.4 $\pm$ 0.7 & 0.4 $\pm$ 0.7\\ 
$a_{0,4}$ &  --  &  --  & 0.0 $\pm$ 1.0\\ 
\hline 
\VE & 40.01 $\pm$ 1.08 & 41.10 $\pm$ 1.14 & 41.10 $\pm$ 1.14\\
\hline 
$\chi^2 / n_\mathrm{df}$ & 24.7/16 & 11.4/16 & 11.3/16\\ 
Prob. & 0.075 & 0.787 & 0.787\\ 
\hline \hline 
\end{tabular} 
\label{table:MIFit_dGdw}
\end{table}
\begin{table}[htb]
\caption{Result of the combined fit to experimental data and different sets of lattice QCD data. The BGL series (Eq.~(\ref{eq:BGL})) is truncated 
after the cubic term.}
\begin{tabular} 
{ @{\hspace{0.1cm}}l@{\hspace{0.1cm}} @{\hspace{0.1cm}}l@{\hspace{0.1cm}} @{\hspace{0.1cm}}r@{\hspace{0.1cm}} @{\hspace{0.1cm}}l@{\hspace{0.1cm}}}
\hline \hline 
Lattice data  & \VE [$10^{-3}$]  & $\chi^2 / n_\mathrm{df}$ & Prob.\\ \hline
FNAL/MILC \cite{lattice2015b}  & $40.96 \pm 1.23$ & 6.01/10 & 0.81\\ 
HPQCD \cite{na2015b} & $41.14 \pm 1.88$ & 4.83/10 & 0.90\\ 
FNAL/MILC \& HPQCD \cite{na2015b,lattice2015b} & $41.10 \pm 1.14$ & 11.35/16 & 0.79\\ 
\hline \hline
\end{tabular} 
\label{table:MIFit_comparison}
\end{table}

\section{Summary}
\label{sec:Summary}

We study the decay {\BDlnu} in 711{\invfb} of Belle $\Upsilon(4S)$~data and reconstruct about 5200 {\Bzerol} and 11,800 {\Bl} decays. We determine the 
differential width {\dGdw} of the decay as a function of the recoil variable $w=V_B\cdot V_D$.

The branching fractions of the decays {\Be}, {\Bmu}, {\Bzeroe}, and {\Bzeromu} are obtained. The isospin-averaged branching fraction 
$\mathcal{B}(\Bzerol)$ is determined to be $(2.31\pm 0.03(\mathrm{stat})\pm 0.11(\mathrm{syst}))\%$.

We interpret our measurement of {\dGdw} in terms of {\VE} by using the currently most established method, {\it i.e.}, by fitting {\dGdw} to the 
Caprini, Lellouch and Neubert (CLN) form-factor parameterization and by dividing {\VGE} by the form factor normalization at zero recoil {\Gone} to 
obtain {\VE}. Assuming the value $\Gone=1.0541\pm 0.0083$~\cite{lattice2015b}, we find $\VE=(40.12\pm 1.34)\times 10^{-3}$. Recent lattice data also 
allows to perform a combined fit to the model-independent form-factor parameterization by Boyd, Grinstein and Lebed (BGL). We find $\VE=(41.10 \pm 
1.14)\times 10^{-3}$ with the lattice QCD data from FNAL/MILC~\cite{lattice2015b} and 
HPQCD~\cite{na2015b}.

Assuming $\etaEW = 1.0066 \pm 0.0016$~\cite{Sirlin1982}, our results correspond to a value of $\Vcb=(39.86\pm 1.33)\times 10^{-3}$ for the fit using the CLN form-factor 
parameterization and \Gone, and $\Vcb=(40.83\pm 1.13)\times 10^{-3}$ for the fit using the BGL parameterization and lattice data.

These results supersede the previous Belle measurement~\cite{Abe:2001yf}.
Compared to the previous analysis by BaBar~\cite{Aubert:2009ac}, we reconstruct about 5 times more {\BDlnu} decays; this results in a significant 
improvement in the precision of the determination of {\VE} from the decay {\BDlnu} to 2.8\%. 
 The value of {\VE} extracted with the combined analysis 
of experimental and LQCD data is in agreement with both {\Vcb} extracted from inclusive semileptonic decays~\cite{Gambino_2013rza} and {\Vcb} from {\BDstarlnu} 
decays~\cite{Dungel:2010uk,Aubert:2007rs}. 
The measured branching fractions are higher although still compatible with those obtained by previous analyses~\cite{Aubert:2009ac}.

\section{Acknowledgments}

We acknowledge useful discussions with Carleton DeTar, Daping Du, Andreas S.\ Kronfeld and Ruth Van de Water from the FNAL/MILC collaboration, and 
with Heechang Na and Junko Shigemitsu from HPQCD, as well as with Paolo Gambino and Florian Bernlochner, and the support of the Mainz Institute for 
Theoretical Physics.

We further thank the KEKB group for the excellent operation of the accelerator; the KEK cryogenics group for the efficient operation of the solenoid; 
and the KEK computer group, the National Institute of Informatics, and the PNNL/EMSL computing group for valuable computing and SINET4 network 
support.  We acknowledge support from the Ministry of Education, Culture, Sports, Science, and Technology (MEXT) of Japan, the Japan Society for the 
Promotion of Science (JSPS), and the Tau-Lepton Physics Research Center of Nagoya University; the Australian Research Council; Austrian Science Fund 
under Grant No.~P 22742-N16 and P 26794-N20; the National Natural Science Foundation of China under Contracts No.~10575109, No.~10775142, 
No.~10875115, No.~11175187, and  No.~11475187; the Chinese Academy of Science Center for Excellence in Particle Physics; the Ministry of Education, 
Youth and Sports of the Czech Republic under Contract No.~LG14034; the Carl Zeiss Foundation, the Deutsche Forschungsgemeinschaft and the 
VolkswagenStiftung; the Department of Science and Technology of India; the Istituto Nazionale di Fisica Nucleare of Italy; the WCU program of the 
Ministry of Education, National Research Foundation (NRF) of Korea Grants No.~2011-0029457,  No.~2012-0008143, No.~2012R1A1A2008330, 
No.~2013R1A1A3007772, No.~2014R1A2A2A01005286, No.~2014R1A2A2A01002734, No.~2015R1A2A2A01003280 , No. 2015H1A2A1033649; the Basic Research Lab 
program under NRF Grant No.~KRF-2011-0020333, Center for Korean J-PARC Users, No.~NRF-2013K1A3A7A06056592; the Brain Korea 21-Plus program and 
Radiation Science Research Institute; the Polish Ministry of Science and Higher Education and the National Science Center; the Ministry of Education 
and Science of the Russian Federation and the Russian Foundation for Basic Research; the Slovenian Research Agency; the Basque Foundation for Science 
(IKERBASQUE) and the Euskal Herriko Unibertsitatea (UPV/EHU) under program UFI 11/55 (Spain); the Swiss National Science Foundation; the National 
Science Council and the Ministry of Education of Taiwan; and the U.S.\ Department of Energy and the National Science Foundation. This work is 
supported by a Grant-in-Aid from MEXT for Science Research in a Priority Area (``New Development of Flavor Physics'') and from JSPS for Creative 
Scientific Research (``Evolution of Tau-lepton Physics'').


\begin{thebibliography}{99}%

\bibitem{cabibbo1963unitary}
N. Cabibbo, Phys. Rev. Lett. \textbf{10}, 531 (1963).

\bibitem {Kobayashi:1973fv}
M. Kobayashi and T. Maskawa, Prog. Theor. Phys. \textbf{49}, 652 (1973).
  
\bibitem{Gambino_2013rza}
P. Gambino and C. Schwanda, Phys. Rev. D \textbf{89}, 014022 (2014), arXiv:1307.4551 [hep-ph].

\bibitem{Dungel:2010uk}
W. Dungel \textit{et al.} (Belle Collaboration), Phys. Rev. D \textbf{82}, 112007 (2010), arXiv:1010.5620 [hep-ex].

\bibitem{Aubert:2007rs}
B. Aubert \textit{et al.} (BaBar Collaboration), Phys. Rev. D \textbf{77}, 032002 (2008), arXiv:0705.4008 [hep-ex].

\bibitem{Aubert:2009ac}
B. Aubert \textit{et al.} (BaBar Collaboration), Phys. Rev. Lett. \textbf{104}, 011802 (2010), arXiv:0904.4063 [hep-ex].

\bibitem{HFAG_averages}
Y. Amhis \textit{et al.} (Heavy Flavor Averaging Group (HFAG)), (2014), arXiv:1412.7515 [hep-ex].  
  
\bibitem{footnote_c}
In all formulae in this paper we set $c = \hbar = 1$.
  
\bibitem{Agashe:2014kda}
K. Olive \textit{et al.} (Particle Data Group), Chin. Phys. C \textbf{38}, 090001 (2014).
  
\bibitem{pbf}
A. Bevan \textit{et al.} (BaBar and Belle Collaborations), Eur. Phys. J. C \textbf{74}, 3026 (2014), arXiv:1406.6311 [hep-ex].  
  
\bibitem{neubert_1991}
M. Neubert, Phys. Lett. B \textbf{264}, 455 (1991).

\bibitem{Sirlin1982}
A. Sirlin, Nucl. Phys. B \textbf{196}, 83 (1982).  

\bibitem{Caprini_et_al}
I. Caprini, L. Lellouch, and M. Neubert, Nucl. Phys. B \textbf{530}, 153 (1998).  
  
\bibitem{boyd1995constraints}
C. G. Boyd, B. Grinstein, and R. F. Lebed, Phys. Rev. Lett. \textbf{74}, 4603 (1995).


\bibitem{lattice2015b}
J. A. Bailey, A. Bazavov, C. Bernard, C. Bouchard, C. DeTar, D. Du, A. X. El-Khadra, J. Foley, E. D. Freeland, \textit{et al.} (Fermilab Lattice and MILC Collaborations), 
Phys. Rev. D \textbf{92}, 034506 (2015), arXiv:1503.07237 [hep-lat].
 
\bibitem{kurokawa2003overview}
S.~Kurokawa and E.~Kikutani, Nucl. Instrum. Methods Phys. Res. Sect.
A {\bf 499}, 1 (2003), and other papers included in this Volume;
T. Abe {\it et al.}, Prog. Theor. Exp. Phys. {\bf 2013}, 03A001 (2013)
and references therein. 
     
\bibitem{hanagaki2002electron}
K. Hanagaki, H. Kakuno, H. Ikeda, T. Iijima, and T. Tsukamoto, Nucl. Instr. and Meth. A \textbf{485}, 490 (2002).  
    
\bibitem{abashian2002muon}
A. Abashian \textit{et al.}, Nucl. Instr. and Meth. A \textbf{491}, 69 (2002).

\bibitem{TheBelleDetector}
A.~Abashian {\it et al.} (Belle Collaboration), Nucl. Instrum. Methods 
Phys. Res. Sect. A {\bf 479}, 117 (2002); also see detector section in
J.Brodzicka {\it et al.}, Prog. Theor. Exp. Phys. {\bf 2012}, 04D001 (2012). 
   
\bibitem{lange2001evtgen}
D. J. Lange, Nucl. Instr. and Meth. A \textbf{462}, 152 (2001).  
  
\bibitem{brun1984geant}
R. Brun \textit{et al.}, GEANT 3.21, Report No, Tech. Rep. (CERN DD/EE/84-1, 1984).  
     
\bibitem{barberio1994photos}
E. Barberio and Z. Wa\c{s}, Comput. Phys. Commun. \textbf{79}, 291 (1994).
   
\bibitem{leibovich1998semileptonic}
A. K. Leibovich, Z. Ligeti, I. W. Stewart, and M. B. Wise, Phys. Rev. D \textbf{57}, 308 (1998).  

\bibitem{abe2001measurement}
K. Abe \textit{et al.}, Phys. Rev. D \textbf{64}, 072001 (2001).  
    
\bibitem{Fox:1978vu}
G. C. Fox and S. Wolfram, Phys. Rev. Lett. \textbf{41}, 1581 (1978).
    
\bibitem{feindt2011hierarchical}%
M. Feindt, F. Keller, M. Kreps, T. Kuhr, S. Neubauer, D. Zander, and A. Zupanc, Nucl. Instr. and Meth. A \textbf{654}, 432 (2011).  
   
\bibitem{footnote_cc}
Charge-conjugate decays are implied throughout this analysis.
   
\bibitem{NeuroBayes}
M. Feindt and U. Kerzel, Nucl. Instr. and Meth. A \textbf{559}, 190 (2006).

\bibitem{BXulnu}
A. Sibidanov \textit{et al.} (Belle Collaboration), Phys. Rev. D \textbf{88}, 032005 (2013), arXiv:1306.2781 [hep-ex].

\bibitem{Fraction_Fit_Barlow}
R. Barlow and C. Beeston, Comput. Phys. Commun. \textbf{77}, 219 (1993).  
   
\bibitem{epaps}
See supplemental material file \texttt{SubsampleResults.csv} in arXiv source files for a table of the measured differential decay widths in the sub-samples and their
full systematic correlation matrix.

\bibitem{na2015b}
H. Na, C. M. Bouchard, G. P. Lepage, C. Monahan, and J. Shigemitsu (HPQCD Collaboration), Phys. Rev. D \textbf{92} 054510 (2015), arXiv:1505.03925 [hep-lat].  

\bibitem{Lepage:1992tx} 
G. P. Lepage, L. Magnea, C. Nakhleh, U. Magnea and K. Hornbostel, Phys. Rev. D \textbf{46} 4052 (1992), arXiv:hep-lat/9205007.
  
\bibitem{ElKhadra:1996mp} 
A. X. El-Khadra, A. S. Kronfeld and P. B. Mackenzie, Phys. Rev. D \textbf{55} 3933 (1997), arXiv:hep-lat/9604004.
    
\bibitem{BCL}
C. Bourrely, L. Lellouch, and I. Caprini, Phys. Rev. D \textbf{79} 013008 (2009), arXiv:0807.2722 [hep-ph].

\bibitem{Abe:2001yf}
K. Abe, \textit{et al.} (Belle Collaboration), Phys. Lett. B \textbf{526}, 258 (2002), arXiv:hep-ex/0111082.


\end{thebibliography}
\end{document}